\documentclass[10pt,journal]{IEEEtran}

\usepackage{graphicx}
\usepackage{amsthm}
\usepackage{amssymb,latexsym,multicol,caption}     
\usepackage{algorithm}  
\usepackage{algorithmic}
\usepackage{epsfig} 
\usepackage{epstopdf}
\usepackage[colorlinks,anchorcolor=red,linkcolor=blue,citecolor=red, urlcolor=black]{hyperref}
\usepackage{color}
\usepackage{multirow}
\usepackage{multicol}
\usepackage{subfigure}
\usepackage{placeins}
\usepackage{bm}
\usepackage{graphics}
\usepackage{footnote}
\usepackage{url}
\usepackage{indentfirst}
\usepackage{longtable}
\usepackage{array}
\usepackage{mdwmath}
\usepackage{mdwtab}
\usepackage{rotating}
\usepackage{color}
\usepackage{morefloats}
\usepackage{slashbox}
\usepackage{cite}
\usepackage{mcite}
\usepackage{eqnarray}
\usepackage{float}
\usepackage{makecell}
\usepackage{amsmath}
\usepackage{threeparttable}
\usepackage{pifont}
\usepackage{cleveref}
\usepackage{balance}
\usepackage{flushend}
\usepackage{makecell}

\graphicspath{fig/fig1}

\renewcommand{\eqref}[1]{(\ref{#1})}

\newcommand{\ml}{\mathcal{L}}
\newcommand{\ms}{\mathcal{S}}
\newcommand{\ma}{\mathcal{A}}
\newcommand{\mc}{\mathcal{C}}
\newcommand{\mo}{\mathcal{O}}
\newcommand{\mb}{\mathcal{B}}
\newcommand{\mz}{\mathcal{Z}}
\newcommand{\mm}{\mathcal{M}}
\newcommand{\sstv}{\operatorname{SSTV}}

\crefname{equation}{Eq.}{Eqs.}
\crefname{figure}{Fig.}{Figs.}
\crefname{table}{Table}{Tables}
\crefname{section}{Section}{Sections}
\crefname{prop}{Proposition}{Propositions}
\crefname{theorem}{Theorem}{Theorems}
\crefname{lemma}{Lemma}{Lemmas}
\crefname{algorithmic}{Algorithm}{Algorithms}

\newtheorem{theorem}{Theorem}[section]
\newtheorem{lemma}{Lemma}[section]

\newcommand{\dd}[1]{{\text{d}}{#1}}

\theoremstyle{plain}

\theoremstyle{definition}

\ifCLASSINFOpdf
\else
\fi

\begin{document}

\title{ Hyperspectral Image Denoising Using Non-convex Local Low-rank and Sparse Separation with Spatial-Spectral Total Variation Regularization }

\author{
Chong Peng, Yang Liu, Yongyong Chen, Xinxin Wu, Andrew Cheng, Zhao Kang, Chenglizhao Chen$^*$, and Qiang Cheng
\thanks{C.P., Y.L., and C.C. are with College of Computer Science and Technology, Qingdao University; Y.C. is with the College of Computer Science, Harbin Institute of Technology;
Z.K. is with the School of Computer Science and Engineering, University of Electronic Science and Technology of China; 
X.W. is with Department of Computer Science, University of Kentucky;
A.C. is with Department of Statistics and Applied Probability, University of California, Santa Barbara;
Q.C. is with the Institute of Biomedical Informatics \& Department of Computer Science, University of Kentucky. }
}


\markboth{}{}
\maketitle

\begin{abstract}

In this paper, we propose a novel nonconvex approach to robust principal component analysis for HSI denoising, 
which focuses on simultaneously developing more accurate approximations to both rank and column-wise sparsity for the low-rank and sparse components, respectively.
In particular, the new method adopts the log-determinant rank approximation and a novel $\ell_{2,\log}$ norm, 
to restrict the local low-rank or column-wisely sparse properties for the component matrices, respectively.
For the $\ell_{2,\log}$-regularized shrinkage problem, we develop an efficient, closed-form solution, which is named $\ell_{2,\log}$-shrinkage operator.
The new regularization and the corresponding operator can be generally used in other problems that require column-wise sparsity. 
Moreover, we impose the spatial-spectral total variation regularization in the log-based nonconvex RPCA model, which enhances the global piece-wise smoothness and spectral consistency from the spatial and spectral views in the recovered HSI.
Extensive experiments on both simulated and real HSIs demonstrate the effectiveness of the proposed method in denoising HSIs.

\end{abstract}


\IEEEpeerreviewmaketitle


\section{Introduction}
Hyperspectral imaging is widely used in various applications, such as biomedical imaging, terrain classification, military surveillance, and remote sensing, etc \cite{tiwari2011An,zhang2015Compression,zhao2014Hyperspectral,Loncan2015hyperspectral}. 
Despite the broad applications of hyperspectral images (HSIs), 
clean HSIs are rarely obtained due to unavoidable corruptions by mixed types of noise, 
such as Gaussian noise, impulse noise, deadlines, and stripes, in the acquisition process \cite{zhang2014hyper}.
Thus, the heavy noise makes it challenging to process HSIs in various applications, such as classification \cite{li2017Hyperspectral} and unmixing \cite{iordeche2011Sparse}.
Due to the adverse effects of noise, 
there is a pressing need for developing effective algorithms to remove noise from HSIs as a pre-processing step of further HSI applications for enhanced learning performance.

HSIs contain hundreds of bands sampled from the visible and infrared range of the electromagnetic spectrum, forming a 3-order tensor structure similar to RBG images with 3 channels. 
In this sense, HSIs can be regarded as extensions of RBG images \cite{he2018hyperspectral}.
HSIs contain three dimensions, including two spatial dimensions (along and across the track) and one spectral dimension (wavelength) \cite{chang2015anisotropic}.
In fact, each band of an HSI can be regarded as a gray-scale image, and existing image denoising algorithms can be readily adopted independently in a band-wise manner to remove noise from HSIs \cite{elad2005image}. 
It should be noted that there is a stark difference between the tasks of denoising single gray-scale images and HSIs. 
Generally, different bands in an HSI usually have high correlations and spectral redundancies, which do not exist in gray-scale images.
Unfortunately, simply applying gray-scale image denoising algorithms to HSI in a band-wise manner ignores the high correlations between different bands,
which omits the special structural information of HSI and thus leads to unsatisfactory denoising performance. 
Thus, despite the great success of gray-scale image denoising algorithms \cite{Dabov2007Image,buades2005a,elad2005image,starck2002the,portilla2003image}, 
there is still an essential demand in designing specialized denoising methods for HSIs due to their particular characteristics.

In the last decades, a number of HSI denoising methods have been developed \cite{zhong2013multiple,othman2006noise,qian2013hyperspectral}.
For example, 
\cite{zhong2013multiple} proposes a unified probabilistic framework, in which the spatial and spectral dependencies are simultaneously adopted;
\cite{qian2013hyperspectral} proposes to take both non-local similarity and spectral-spatial structure of HSIs into consideration in a sparse representation-based framework. 
Moreover, methods such as principal component analysis (PCA), wavelet shrinkage, anisotropic diffusion, multitask sparse matrix factorization, and tensor decomposition, etc.,
have been considered for HSI denoising \cite{ye2015multitask,chen2011denoising,duarte2007comparative,wang2010anisotropic,xu2017non,chen2020nonlocal,zheng2020double}.
Most of the above methods require some specific prior knowledge of the noise.
Unfortunately, such knowledge is rarely available and generally limited in real-world application, and thus the above methods cannot remove all types of noise from HSIs.
Hence, more effective methods are urgently in demand. 

{
Recently, low-rank techniques have been developed and successful in various applications such as subspace clustering and HSI denoising \cite{wang2021tensor,wei2019intracluster,cao2019hyperspectral}.
For HSI denoising, various approaches have been attempted, such as low-rank matrix factorization \cite{du2018bandwise,xu2017non}, low-rank tensor decomposition \cite{zhang2019hyperspectral,fan2018spatial}, low-rank dictionary learning \cite{gong2020a}, low-rank regularization methods \cite{xue2017joint}, etc.
}
%
The success of these methods in HSI denoising is based on a natural assumption that a scene of a clean HSI is composed of much fewer endmembers than spectral bands and pixels, 
which reveals the natural low-rank structures of HSI \cite{bioucas2011an,he2016sparsity}. 
Thus, low-rank based techniques have been widely adopted to remove noise from HSIs,  
among which PCA \cite{chien1999interference} and low-rank matrix factorization based methods \cite{Xu2017Denoising} are typical ones.
Unfortunately, these methods are known to suffer from being sensitive to outliers that commonly exist in HSIs \cite{Cand2011Robust}.
Thus, it is generally difficult for these methods to completely remove outliers, such as stripes, deadlines, and impulse noise from HSIs.
To combat this issue, RPCA models the outliers by separating a sparse component, which significantly improves the robustness.
In the original paper of RPCA \cite{Cand2011Robust}, 
it is proved that under certain conditions there is a high probability to correctly separate the low-rank and sparse components from the observed data.

Traditional low-rank matrix recovery models such as RPCA adopt the nuclear norm to restrict the low-rank property of the target matrix.
Recently, it has been pointed out that the nuclear norm as used in the RPCA model is not accurate in approximating the rank function \cite{peng2015subspace,xie2019hyperspectral,chen2017denoising,hu2013fast},
which may lead to degraded performance in low-rank recovery \cite{peng2015subspace,peng2020robust}.
To solve this issue, nonconvex approaches are attempted to better approximate the rank function in the RPCA framework \cite{peng2020robust}, which achieves promising performance.
{Despite the success of nonconvex rank approximations, nonconvex approximations of the sparsity are rarely considered in RPCA model. 
In this paper, we point out that there is a close connection between the nuclear norm and the $\ell_{2,1}$ norm, i.e., the rank and column-wise sparsity approximations.
For nonconvex rank approximations, the improved approximating behaviors actually benefit from the improved approximation to the sparsity of the singular values,
which inspires us to improve the approximating behavior of the column-wise sparsity in RPCA framework with nonconvex approach for more accurate low-rank and sparse components separation in HSI denoising \cite{xie2020hyperspectral}.  
In particular, we propose a log-based column-wisely sparse approximation, namely the $\ell_{2,\log}$ norm, which admits some favorable properties, which will be discussed with details in later sections.
}

{While low-rank recovery models are effective in HSI denoising, 
they only explore the correlation between spectral bands of HSIs with the low-rank constraints while omitting spatial correlation of local neighboring pixels.}
Several approaches have been attempted in low-rank models to incorporate spatial information of HSIs, such as wavelets \cite{chen2011denoising,rasti2014wavelet}, total variation (TV) regularization \cite{he2015total,wang2018low}, and sparse representation \cite{zhao2015hyperspectral,zhuang2018fast}, etc. 
These methods follow a common strategy by restricting low-rank structure on the overall HSI with specialized regularization constraints imposed on it.
However, the same material from different local areas of HSI may have starkly different spectral signatures, 
which leads to the increased rank of the overall HSI. 
Meanwhile, local areas are likely to contain the same material and thus the same spectral signature, which implies local-low rank property of HSI.
This inspires the segmentation of the HSI into overlapping 3D patches \cite{zhang2013hyperspectral,xie2016hyperspectral}, where they are processed sequentially with the RPCA model.
Oftentimes, the sparse noise exists in the same location of some bands and naturally forms a local low-rank structure.
Such noise is considered as structured sparse noise and is often mathematically treated as part of the low-rank component by spectral low-rank property.
Consequently, it is difficult for local low-rank models to remove such noise only with spectral correlation of HSI and it is demanding to exploit spatial constraint for improved denoising performance.
In fact, clean HSI favors global piece-wise smoothness and spectral consistency from the spatial and spectral viewpoints, respectively, which are destroyed by the structured sparse noise.
Thus, in this paper, we follow \cite{he2018hyperspectral} and simultaneously seek the separation of low-rank and sparse matrices with spatial-spectral TV (SSTV) for enhanced spatial-spectral piece-wise smoothness and consistency. 
It is noted that the log-based non-convex approximations to the rank and column-wise sparsity ensure that the low-rank and sparse components can be more accurately separated than traditional methods. Meanwhile, the SSTV term enhances the global piece-wise smoothness and spectral consistency of the recovered HSI to remove potential noise remaining in the low-rank component.

We summarize the key contributions of our paper as follows:
{1) We propose a novel nonconvex RPCA model with simultaneous log-based non-convex approximations to the rank and column-wise sparsity, respectively.
The proposed $\ell_{2,\log}$ norm is more accurate than the widely used $\ell_{2,1}$ norm.
2) For the $\ell_{2,\log}$-norm regularized shrinkage problem, we formally provide a closed-form solution, which is efficient and can be generally used in various problems that restrict column-wise sparsity.
3) The SSTV is first integrated with log-based non-convex RPCA model for simultaneous more accurate local low-rank and sparse separation and enhanced global piece-wise smoothness as well as spectral consistency. 
4) Elegant theoretical analysis is provided for the proposed optimization algorithm}. 
5) Superior performance is observed compared with state-of-the-art baseline methods, which confirms the effectiveness of the proposed method. 


\section{Related Work}
\label{sec_related}
In this section, we will briefly review a few techniques that are closely related with our work.

\subsection{Robust Principal Component Analysis}
Given data matrix $X$, RPCA assumes that the data can be decomposed into a low-rank $L$ and a sparse $S$, which can be mathematically formed as $X=L+S$.
To obtain the two components, the classic RPCA aims at solving the following constrained optimization problem \cite{Cand2011Robust}:
\begin{equation}
\label{eq_rpca_l1}
\min_{L,S} \|L\|_* + \lambda \|S\|_{1},	\quad s.t.\quad X = L+S,
\end{equation}
where $\|\cdot\|_*$ is the nuclear norm that adds all singular values of the input matrix, $\|\cdot\|_1$ is the $\ell_1$ norm that adds the absolute values of all elements of the input matrix, and $\lambda\ge 0$ is a balancing parameter.

\subsection{Spatial-Spectral Total Variation}
For HSI, it is natural that two nearby bands are very similar, which indicates spectral consistency. 
Also, HSIs have spatial correlations.
Thus, it is convincing to adopt TV norm from both spatial and spectral directions for HSI.
For an observed tensor cube $\mathcal{M}$, we denote its $(i,j,b)$-th element by $(\mathcal{M})_{i,j,b}$, 
where $i$ and $j$ represent the horizontal and vertical directions while $b$ corresponds to the spectral direction, respectively. 
Then the anisotropic spatial-spectral TV norm can be formulated as 
\begin{equation}
\label{eq_sstv1}
\|\mathcal{M}\|_{\sstv} = \|\textbf{D}_x \mm\|_1 + \|\textbf{D}_y \mm\|_1 + \|\textbf{D}_z \mm\|_1,
\end{equation}
where $\textbf{D}_x$, $\textbf{D}_y$, and $\textbf{D}_z$ perform first-order discrete differences of $\mm$ in horizontal, vertical, and spectral directions, respectively, which are defined as 
\begin{eqnarray}
\begin{cases}
& \textbf{D}_x \mm = \textbf{vec}\Big(\big[ (\mm)_{i+1,j,b} - (\mm)_{i,j,b} \big]\Big)	\\
& \textbf{D}_y \mm = \textbf{vec}\Big(\big[ (\mm)_{i,j+1,b} - (\mm)_{i,j,b} \big]\Big)	\\
& \textbf{D}_z \mm = \textbf{vec}\Big(\big[ (\mm)_{i,j,b+1} - (\mm)_{i,j,b} \big]\Big)
\end{cases}
\end{eqnarray}
with periodic boundary conditions. 
Here, $\textbf{vec}(\cdot)$ is a linear operator that reshapes a tensor into a vector. 
With \cref{eq_sstv1}, the piece-wise smoothness is restricted in both the spatial and spectral directions.
It is pointed out that the equal weights for the gradients along different dimensions might not be proper and \cref{eq_sstv1} is further extended to the following anisotropic SSTV regularization:
\begin{equation}
\label{eq_sstv}
\|\mathcal{M}\|_{\sstv} = \tau_x \|\textbf{D}_x \mm\|_1 + \tau_y\|\textbf{D}_y \mm\|_1 + \tau_z\|\textbf{D}_z \mm\|_1,
\end{equation}
where $\tau_x$, $\tau_y$, and $\tau_z$ are balancing parameters. In this paper, we set $[\tau_x,\tau_y,\tau_z] = [1,1,0.5]$ as recommended in \cite{chan2011an,he2018hyperspectral}.

\section{SSTV Regularized Nonconvex RPCA}
\label{sec_proposed}
For the observed HSI $\mo\in\mathcal{R}^{M\times N\times p}$, 
it is natural to separate the clean part $\ml\in\mathcal{R}^{M\times N\times p}$ and noise part $\ms\in\mathcal{R}^{M\times N\times p}$ as $\mo = \ml + \ms.$
Due to the nature of HSIs, usually the adjacent bands are highly correlated, which leads to the low-rank structure. 
%
For HSIs, the spectral signatures of the same local area are more likely to be the same, which inspires us to exploit the local low-rank property of the HSIs.
Thus, we divide the HSI into overlapping patches and exploit the patch-wise local low-rank structure. 
%
Specifically, for the tensor $\ml$, we first find an $m\times n\times p$ patch cube centralized at location $(i,j)$.
Then we vectorize all patch bands and form its corresponding Casorati matrix $\ml_{i,j}\in\mathcal{R}^{mn\times p}$, with each column of $\ml_{i,j}$ being a vectorized patch band.
Similarly, we define the Casorati matrices $\mo_{i,j}$ and $\ms_{i,j}$ from $\mo$ and $\ms$. 
{
With these definitions, the target is to separate low-rank $\ml_{i,j}$ and sparse $\ms_{i,j}$ from the observed matrix $\mo_{i,j}$.
To keep the spatial structural information of $\ms_{i,j}$, we adopt the $\ell_{2,1}$ norm in the low-rank and sparse separation model \cite{Cand2011Robust,kang2015robust}, which leads to the following patch-based RPCA model \cite{he2018hyperspectral}: 
\begin{equation}
\label{eq_obj_nuclear_21}
\begin{aligned}
&	\min_{\ml,\ms} \sum_{i}\sum_{j} \bigg\{ \|\ml_{i,j}\|_* + \lambda \|\ms_{i,j}\|_{2,1} \bigg\}	\\
&	 s.t. \quad \mathcal{O}_{i,j} = \ml_{i,j} + \ms_{i,j},
\end{aligned}
\end{equation}
where $\|\ms_{i,j}\|_{2,1} = \sum_{t=1}^{p} \|(\ms_{i,j})_t\|_2$ is the $\ell_{2,1}$ norm that restricts column-wisely sparse structure for $\ms_{i,j}$ with $\|\cdot\|_2$ being the $\ell_2$ norm, 
and $i,j$ vary such that the patches overlap and cover the overall data.
In this paper, we follow the strategy in \cite{he2018hyperspectral} and simultaneously process all patches such that the correlation of the patches can be preserved.
}

Recently, it is pointed out that the nuclear norm is not accurate in approximating the true rank of a matrix,
and the use of the nuclear norm may lead to degraded performance in low-rank recovery problems \cite{peng2020robust,peng2015subspace}.
{It is natural that the rank should be estimated properly in low-rank recovery problems.
To combat the issue of the nuclear norm, nonconvex approaches to rank approximation have been developed with more accurate approximating behavior and have drawn significant attentions in various problems with promising performance \cite{xie2016hyperspectral,xie2020hyperspectral,hu2013fast}.}
In this paper, due to the efficiency in optimization, we adopt the log-determinant rank approximation \cite{peng2020robust,peng2015subspace} to replace the nuclear norm in above model.
The log-determinant rank approximation is defined as 
\begin{equation}
\begin{aligned}
 \|\ml_{i,j}\|_{\log\det} & = \log\det (I + (\ml_{i,j}^T \ml_{i,j})^{\frac{1}{2}} ) \\
&  = \sum\nolimits_{s = 1}^{p} \log (1+ \sigma_s  (\ml_{i,j}) ),
\end{aligned}
\end{equation}
where $I$ is an identity matrix with proper size and $\sigma_s(\cdot)$ returns the $s$th largest singular value of the input matrix.
Thus, model \cref{eq_obj_nuclear_21} is further developed into 
\begin{equation}
\label{eq_obj_logdet_21}
\begin{aligned}
&	\min_{\ml,\ms} \sum_{i}\sum_{j} \Big\{ \log\det (I + (\ml_{i,j}^T \ml_{i,j})^{\frac{1}{2}} ) + \lambda \|\ms_{i,j}\|_{2,1} \Big\}	\\
&	 s.t. \quad \mathcal{O}_{i,j} = \ml_{i,j} + \ms_{i,j}.
\end{aligned}
\end{equation}
\begin{figure}[h]
	\centering
	{\includegraphics[width=1\columnwidth]{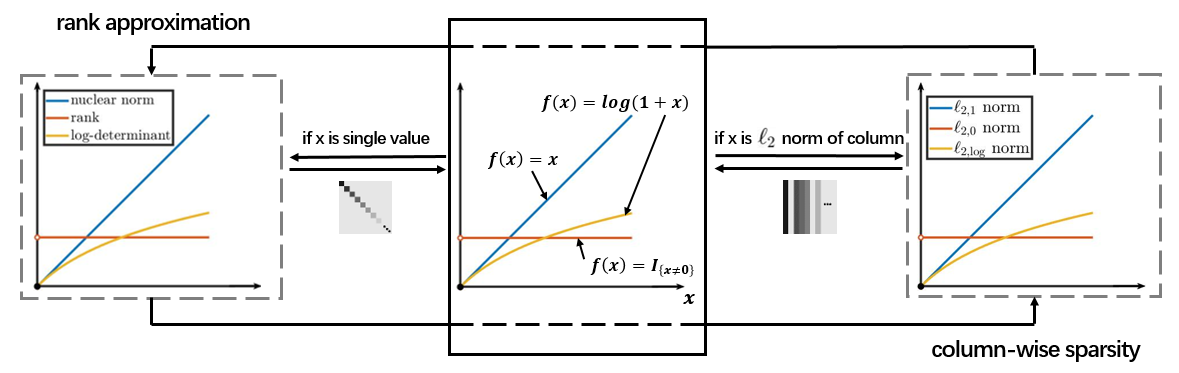} }
	\caption{ Illustration of the relationship between the log-determinant rank approximation and the column-wisely sparse approximation, i.e., the $\ell_{2,\log}$ norm. }
	\label{fig_relation}
\end{figure}
%
In fact, the success of nonconvex approach to rank approximation is based on the closer approximation to the sparseness of the singular values than the nuclear norm. 
{As shown in \cref{fig_relation}, there is a close connection between the low-rank and sparse approximation problems.
It is seen that the nuclear and the $\ell_{2,1}$ norms are essentially the $\ell_1$ norm of singular values and column-wise $\ell_2$ norms, respectively. 
Thus, by adding the $\ell_2$ norm of all columns, the $\ell_{2,1}$ norm suffers from a similar issue to the nuclear norm and thus is not accurate in approximating the true column-wise sparsity.
Inspired by the success of log-determinant rank approximation, in this paper we propose a log-based column-wisely sparse approximation, the $\ell_{2,\log}$ (pseudo) norm, to restrict the column-wise sparseness of $\ms_{ij}$:}
\begin{equation}
\label{eq_L2log}
\|A\|_{2,\log} = \sum_{j}\log (1+ \|a_j\|_2),
\end{equation}
where $A=[\cdots,a_j,\cdots]$ is the input matrix, and $\|\cdot\|_2$ is the $\ell_2$ norm. 
The $\ell_{2,\log}$ norm admits the following properties:
\begin{itemize}
\item[1)] $\|A\|_{2,\log}\ge 0$ for any matrix $A$ and $\|A\|_{2,\log} = 0$ if and only if $A=0$. 
\item[2)] The $\ell_{2,\log}$ norm is nonconvex, continuous, and differentiable, 
where $\frac{\partial \|A\|_{2,\log}}{\partial A} = [ \cdots,\frac{a_j}{\|a_j\|_2(1+\|a_j\|_2)},\cdots ]$.
{
\item[3)] For $\|a_j\|_2 \not= 0$, ${\sum_j\log (1+ \|a_j\|_2)}<{\sum_j \|a_j\|_2}$, which implies that $\ell_{2,\log}$ norm is more accurate than the $\ell_{2,1}$ norm in approximating the sparsity for large values and noise effects for small values, respectively.
\item[4)] As will be clear in later section, the $\ell_{2,\log}$ norm has smaller expectation than $\ell_2$ norm and is better in estimating noise effects.
\item[5)] As will be clear in later section, the triangle inequality holds for $\ell_{2,\log}$.}
\end{itemize}
Thus, with the $\ell_{2,\log}$ norm, \cref{eq_obj_logdet_21} is further developed into 
%
\begin{equation}
\label{eq_obj_log}
\begin{aligned}
&	\min_{\ml,\ms} \sum_{i,j} \log\det (I + (\ml_{i,j}^T \ml_{i,j})^{\frac{1}{2}})  \\
&	\qquad\qquad\qquad + \lambda  \sum_{i,j} \sum_{t} \log (1+ \|(\ms_{i,j})_t\|_2 )	\\
&	 s.t. \quad \mathcal{O}_{i,j} = \ml_{i,j} + \ms_{i,j}.
\end{aligned}
\end{equation}
%
%
It is seen that \cref{eq_obj_log} is a patch-based model, which learns the local low-rank property of HSIs.
{
Although patch-based low-rank recovery methods are successful in HSI denoising \cite{dong2015low,xie2016hyperspectral,zhang2014hyper},
they fail to exploit global structural information, i.e., correlations of spatial pixels and spectral bands of HSIs, which may lead to failure in noise removal \cite{he2018hyperspectral}.
To address this issue, we integrate the SSTV with \cref{eq_obj_log} to seek local low-rank and TV properties in both spatial and spectral domains and obtain the following model:
}
\begin{equation}
\label{eq_obj}
\begin{aligned}
&	\min_{\ml,\ms} \sum_{i,j} \bigg \{\|\ml_{i,j}\|_{\log\det} + \lambda \|\ms_{i,j}\|_{2,\log} \bigg \}  + \gamma \|\ml\|_{\sstv} \\
&	 s.t. \quad \mathcal{O}_{i,j} = \ml_{i,j} + \ms_{i,j},
\end{aligned}
\end{equation} 
where $\gamma \ge 0$ is a balancing parameter. 
We name the model in \cref{eq_obj} the Log-based Local Low-rank and Sparse separation model with Spatial-Spectral Total Variation (L$^3$S$^3$TV). 
It is seen that the L$^3$S$^3$TV model exploits both local and global structures of HSIs, where the local low-rank property of HSIs are sought with the first two terms, and the correlations of spatial pixels and spectral bands are exploited with the SSTV term.
For the optimization, we will develop an efficient algorithm in the next section.

{
	{\sl{Remark:}}
	For the $\ell_{2,\log}$ norm, we have the following conclusions. 
	The following discussions can be generalized to the other columns and the overall matrix.
	For a specific column of a matrix, we denote the elements by $X_1,\cdots,X_d.$
	Then, for any distributions of $X_i's$, the expectation of the log-based approximation is generally less than the $\ell_2$-based approximation, which can be formally analyzed in the following:
	%
	\begin{equation}
	\begin{aligned}
	& \mathbf{E} \Bigg(\log \Bigg(1 + \sqrt{\sum\nolimits_{i=1}^d X_{i}^2}\Bigg) \Bigg) \\
	= &  \int_{0}^{+\infty} \log  (1 + \sqrt{ y } ) f_{\sum_{i=1}^d X_{i}^2}(y) \dd{y} \\  
	< & \int_{0}^{+\infty} \sqrt{y} f_{\sum_{i=1}^d X_{i}^2}(y)  \dd{y} = \mathbf{E} \Bigg(\sqrt{\sum\nolimits_{i=1}^d X_{i}^2} \quad \Bigg), \\
	\end{aligned}
	\end{equation}
where $f_{\sum_{i=1}^d X_{i}^2}(y)$ is the probability density function for $y=\sum_{i=1}^d X_{i}^2$.
Specially, regarding two types of most widely considered distributions, i.e., normal and uniform distributions, we derive the bounds information in the Appendix.

	Moreover, for essentially small values of $\sum_{i=1}^d X_{i}^2$, it is natural that such values correspond to noise and the corresponding columns are indeed sparse.
	Thus, for such small values, it is essentially important that the approximation is close to 0 rather than 1 to distinguish noise effects and useful information.
	It is noted that $\log(1+\sqrt{x})<\sqrt{x}$ holds for small $x$,
	which indicates that the log-based approximation is closer to 0 than the $\ell_2$-based approach and thus is more accurate in approximating the real sparsity. 
	Thus, it is expected that the log-based approximation is more accurate in approximating the real sparse indicator of the columns than the $\ell_2$-based approach.
	
}

\section{Optimization}
\label{sec_optimization}
In this section, we will develop an efficient optimization algorithm for \cref{eq_obj} based on the augmented Lagrange multiplier method (ALM).
In particular, we first introduce some auxiliary variables to \cref{eq_obj} and obtain the following equivalent model:
\begin{equation}
\begin{aligned}
& \min_{\ml,\ms,\ma,\mb,\mc} \sum_{i,j} \bigg\{\|\ml_{i,j}\|_{\log\det} + \lambda \|\ms_{i,j}\|_{2,\log} \bigg\} + \gamma \|\ml\|_{\sstv} \\
& s.t. \quad   \mathcal{O}_{i,j} = \ml_{i,j} + \ms_{i,j}, \ml_{i,j} = \ma_{i,j}, \ma = \mb, \mc = \textbf{D}\mb,
\end{aligned}
\end{equation} 
where $\textbf{D} = [\tau_x\textbf{D}_x, \tau_y\textbf{D}_y, \tau_z\textbf{D}_z]$ denotes the TV operator in the spatial and spectral directions and 
\begin{equation}
\textbf{D}\mb = [ \tau_x\textbf{D}_x \mb, \tau_y\textbf{D}_y \mb, \tau_z\textbf{D}_z \mb ] \in \mathcal{R}^{MNp\times 3}.
\end{equation}
Then we need to optimize the augmented Lagrange function as follows:
\begin{equation}
\begin{aligned}
& \min_{\ml,\ms,\ma,\mb,\mc,\mz^{\ma},\mz^{\mb},\mz^{\mc},\mz^{\mo}} \sum_{i,j} \bigg\{\|\ml_{i,j}\|_{\log\det} \!+\! \lambda \|\ms_{i,j}\|_{2,\log}  \\
& \qquad\qquad + \frac{\rho}{2} \| \mo_{i,j} - \ml_{i,j} - \ms_{i,j} + \frac{1}{\rho}{\mz}^{\mo}_{i,j} \|_F^2	\\
& \qquad\qquad + \frac{\rho}{2} \| \ml_{i,j} - \ma_{i,j} + \frac{1}{\rho}{\mz}^{\ma}_{i,j} \|_F^2 \bigg\}	\\
& \qquad\qquad + \gamma \|\mc\|_{1} + \frac{\rho}{2} \| \mc - \textbf{D}\mb + \frac{1}{\rho}{\mz}^{\mc} \|_F^2	\\
& \qquad\qquad + \frac{\rho}{2} \| \ma - \mb + \frac{1}{\rho}{\mz}^{\mb} \|_F^2,
\end{aligned}
\end{equation} 
where $\mo,\ml,\ms,\ma,\mb,\mz^{\mo},\mz^{\ma},\mz^{\mb}\in\mathcal{R}^{M\times N\times p}$, $\mc,\mz^{\mc}\in\mathcal{R}^{MNp\times 3}$,
and $\|\mm\|_F$ denotes $\sqrt{\sum_{a,b,c}(\mm)_{a,b,c}^2}$ for tensor $\mm$ for ease of notation.
{Next, we will develop the alternating optimization strategies for each variable, respectively. }

\subsection{$\ml_{i,j}$-minimization}
\label{sec_opt_L}
To optimize $\ml_{i,j}$, we have the following sub-problem
\begin{equation}
\label{eq_Lij}
\begin{aligned}
& \min_{\ml_{i,j}} \sum_{i,j} \bigg\{\|\ml_{i,j}\|_{\log\det} + \frac{\rho}{2} \| \ml_{i,j} - \ma_{i,j} + \frac{1}{\rho}{\mz}^{\ma}_{i,j} \|_F^2	\\
& \qquad\qquad\qquad 	+ \frac{\rho}{2} \| \mo_{i,j} - \ml_{i,j} - \ms_{i,j} + \frac{1}{\rho}{\mz}^{\mo}_{i,j} \|_F^2 \bigg\}.\\
\end{aligned}
\end{equation} 
The above problem can be solved for each $\ml_{i,j}$ independently with 
\begin{equation}
\label{eq_Lij_each}
\begin{aligned}
& \min_{\ml_{i,j}} \|\ml_{i,j}\|_{\log\det} + \frac{\rho}{2} \| \ml_{i,j} - \ma_{i,j} + \frac{1}{\rho}{\mz}^{\ma}_{i,j} \|_F^2	\\
& \qquad\qquad\qquad 	+ \frac{\rho}{2} \| \mo_{i,j} - \ml_{i,j} - \ms_{i,j} + \frac{1}{\rho}{\mz}^{\mo}_{i,j} \|_F^2.
\end{aligned}
\end{equation} 
Let $\mathcal{X}_{ij} = \mo_{i,j} -  \ms_{i,j} + \frac{1}{\rho}{\mz}^{\mo}_{i,j}  + \ma_{i,j} - \frac{1}{\rho}{\mz}^{\ma}_{i,j}$,
then with straightforward algebra, the problem \cref{eq_Lij_each} is equivalent to 
\begin{equation}
\label{eq_sub_L}
\begin{aligned}
& \min_{\ml_{i,j}} \frac{1}{\rho}\|\ml_{i,j}\|_{\log\det} + \frac{1}{2} \| \ml_{i,j} - \mathcal{X}_{i,j} \|_F^2.
\end{aligned}
\end{equation}
For a matrix $D$, we define $\mathcal{P}(D)$, $\mathcal{Q}(D)$, and $\sigma_i(D)$ to be its left and right singular vectors and the $i$-th largest singular value, respectively.
Then, similar to \cite{peng2015subspace,peng2020robust}, 
\cref{eq_sub_L} admits a closed-form solution with the following operator:
\begin{equation}
\label{eq_sol_L}
\mathcal{L}_{i,j} = \mathcal{D}_{\frac{1}{\rho}}(\mathcal{X}_{i,j}),
\end{equation}
where $\mathcal{D}_{ \delta }(D) = \mathcal{P}(D) \text{diag}\{\sigma_i^*\} (\mathcal{Q}(D))^T$, with 
\begin{eqnarray}
\sigma_{i}^* = 
\begin{cases}
\xi,&\mbox{ if $f_i(\xi) \le f_i(0)$ and $ (1 + \sigma_{i}(D))^2 > 4\delta$, }	\\
0, 	&\mbox{ otherwise, }
\end{cases}
\end{eqnarray}
%
where $$f_i(x) = \frac{1}{2}(x-\sigma_i(D))^2 + \tau \log (1+x)$$ and $$\xi = \frac{\sigma_i(D)-1}{2} + \sqrt{\frac{(1+\sigma_i(D))^2}{4} - \delta }.$$

\subsection{$\ms_{i,j}$-minimization}
The sub-problem associated with optimization of $\ms_{i,j}$ is 
\begin{equation}
\label{eq_sub_S}
\begin{aligned}
& \min_{\ms_{i,j}}  \sum_{i,j} \bigg \{ \lambda \|\ms_{i,j}\|_{2,\log} \\
& \qquad\qquad + \frac{\rho}{2} \| \mo_{i,j} - \ml_{i,j} - \ms_{i,j} + \frac{1}{\rho}{\mz}^{\mo}_{i,j} \|_F^2 \bigg \},
\end{aligned}
\end{equation} 
which can be solved in an element-wise manner with 
\begin{equation}
\label{eq_sub_S_elementwise}
\min_{\ms_{i,j}} \frac{\lambda}{\rho} \|\ms_{i,j}\|_{2,\log} + \frac{1}{2} \| \mo_{i,j} - \ml_{i,j} - \ms_{i,j} + \frac{1}{\rho}{\mz}^{\mo}_{i,j} \|_F^2.
\end{equation} 
Problems in a format of \cref{eq_sub_S_elementwise} is a $\ell_{2,\log}$-regularized shrinkage problem.
We formally have the following theorem to solve it. 
\begin{theorem}[$\ell_{2,\log}$-shrinkage operator]
\label{thm_2log}
	
	Given matrix $Y\in\mathcal{R}^{d\times n}$ and a nonnegative parameter $\alpha$, the following problem
	\begin{equation}
	\label{eq_soft_thres}
	\min_{W \in \mathcal{R}^{d\times n}} \frac{1}{2} \|Y-W\|_F^2 + \alpha \|W\|_{2,\log}
	\end{equation}
	%
	is called $\ell_{2,\log}$-regularized shrinkage problem, which admits closed-form solution in a column-wise manner:
	\begin{equation}
	\label{eq_sol_soft_thres}
	\!w_i \!=\! 
	\begin{cases}
	\frac{\xi}{\|y_i\|_2} y_i, & \mbox{ if $f_i(\xi) \le \frac{\|y_i\|_2^2}{2}, 
		\frac{(1 \!+\! \|y_i\|_2)^2}{4}\!>\!\alpha$, and $\xi >0$}	\\
	0, 	& \!\mbox{ otherwise, }
	\end{cases}
	\end{equation}
	where $f_i(x) = \frac{1}{2}(x-\|y_i\|_2)^2 + \alpha \log (1+x),$ and $\xi = \frac{\|y_i\|_2-1}{2} + \sqrt{\frac{(1+\|y_i\|_2)^2}{4} - \alpha }.$
	
\end{theorem}

\noindent{{\em{Proof}}}:
	It is easy to see that \cref{eq_soft_thres} can be solved with respect to each $w_i$ independently.
	For $w_i$, the sub-problem is $$	\min_{w_i} \frac{1}{2} \|y_i-w_i\|_2^2 + \alpha \operatorname{log}( 1+\|w_i\|_2 ).$$
	We may treat $w_i$ as a special matrix and perform thin SVD to it. 
	Then it is seen that $w_i$ has exactly one singular value, which is $\operatorname{\sigma}(w_i) = \sqrt{w_i^Tw_i} = \|w_i\|_2$, 
	where $\operatorname{\sigma}(\cdot)$ is the singular value of the input vector.
	Thus, optimizing $w_i$ is equivalent to 
	\begin{equation}
	\label{eq_soft_thres_wi_svd}
	\min_{w_i} \frac{1}{2} \|y_i-w_i\|_2^2 + \alpha \operatorname{log}( 1+\sigma(w_i) ).
	\end{equation}
	Hence, according to \cite{peng2015subspace,peng2020robust} and \cref{sec_opt_L}, the solution to \cref{eq_soft_thres_wi_svd} is obtained with $w_i = u_i \sigma^*(w_i) v_i^T,$
	where $u_i$ and $v_i$ are left and right singular vectors of $y_i$, respectively, and 
	\begin{eqnarray}
	\!\!\sigma^*(w_i)\!\! =\!\! 
	\begin{cases}
	\!\xi ,& \! \mbox{ if $f_i(\xi) \!\le\! f_i(0), \frac{(1 \!+\! \sigma(y_i))^2}{4} \!>\! \alpha$, and $  \xi>0$ }	\\
	\! 0, 	& \! \mbox{ otherwise, }
	\end{cases}
	\end{eqnarray}
	with 
$$f_i(x) = \frac{1}{4}(x - \sigma(y_i))^2 + \alpha \log (1 + x),$$ and $$\xi =\frac{\sigma(y_i)-1}{2} + \sqrt{\frac{(1+\sigma(y_i))^2}{4} - \alpha }.$$
{If $y_i = 0$, then $w_i = 0$ is clearly the optimal solution and \cref{eq_sol_soft_thres} is true. Thus, it suffices to consider $y_i \ne 0$. 
	In this case, it is straightforward that $y_i = \frac{y_i}{\|y_i\|_2}\|y_i\|_2 [1]$ is a thin SVD of $y_i$.
	Here, the notation $[1]$ represents a special row matrix with only one column that is $1$. }
	We substitute $u_i = \frac{y_i}{\|y_i\|_2}$, $\sigma(y_i) = \|y_i\|_2$, and $v_i = [1]$ into above equations, 
	which leads to \cref{eq_sol_soft_thres} and concludes the proof. $\hfill \Box$


For ease of notation, we denote the $\ell_{2,\log}$-shrinkage operator of \cref{eq_sol_soft_thres} as $\mathcal{T}_{\alpha}(Y)$, 
generating the solution to \cref{eq_sub_S}:
\begin{equation}
\label{eq_sol_S}
\ms_{i,j} = \mathcal{T}_{\frac{\lambda}{\rho}}( \mo_{i,j} - \ml_{i,j} + {\mz}^{\mo}_{i,j} / \rho ).
\end{equation}
{
It is seen that the log-based shrinkage problem admits closed-form solution, which is more efficient than existing approaches that adopt iterative optimization strategy such as \cite{xie2020hyperspectral,chen2017denoising}. 
	
}

\subsection{$\ma$-minimization}
The sub-problem associated with optimization of $\ma$ is 
\begin{equation}
\label{eq_sub_A}
\begin{aligned}
\!\min_{\ma} \! \sum_{i,j} \! \frac{\rho}{2} \| \ml_{i,j} \!-\! \ma_{i,j} \!+\! \frac{1}{\rho}{\mz}^{\ma}_{i,j} \|_F^2 
\!+\!  \frac{\rho}{2} \| \ma - \mb + \frac{1}{\rho}{\mz}^{\mb} \|_F^2 \\
\end{aligned}
\end{equation} 
%
We define $\textbf{1}_{\{\cdot\}}$ to be an indicator function, which returns 1 if the condition in the subscript is satisfied and 0 otherwise.
Then the above problem for $\ma$ can be rewritten in an element-wise manner for each $(\ma)_{a,b,c}$ as follows:
\begin{equation}
\begin{aligned}
& \arg\min_{(\ma)_{a,b,c}} \bigg((\ma)_{a,b,c} - (\mb)_{a,b,c} + \frac{1}{\rho}({\mz}^{\mb})_{a,b,c} \bigg)^2  \\
&  + \bigg( (\ml)_{a,b,c} - (\ma)_{a,b,c} + \frac{1}{\rho}({\mz}^{\ma})_{a,b,c} \bigg)^2 \sum_{i,j} \textbf{1}_{\{(\ma)_{a,b,c}\in\ma_{i,j}\}}, 	\\ 
\end{aligned}
\end{equation} 
where $\sum_{i,j} \textbf{1}_{\{(\ma)_{a,b,c}\in\ma_{i,j}\}}$ counts the number of times that $(\ma)_{a,b,c}$ is overlapped in \cref{eq_sub_A}. 
It is seen that the problem is quadratic in $(\ma)_{a,b,c}$, which admits closed-form solution with the following first-order optimality condition.
Thus, we have the following closed-form solution for $\ma$.
\begin{equation}
\label{eq_sol_A}
\begin{aligned}
& (\ma)_{a,b,c} \\
= & \frac{1}{ 1+\sum_{i,j} \textbf{1}_{\{(\ma)_{a,b,c}\in\ma_{i,j}\}} }\bigg\{(\mb)_{a,b,c} - \frac{1}{\rho}({\mz}^{\mb})_{a,b,c} \\
&\space	+ \sum_{i,j} \textbf{1}_{\{(\ma)_{a,b,c}\in\ma_{i,j}\}} \bigg((\ml)_{a,b,c} + \frac{1}{\rho}({\mz}^{\ma})_{a,b,c} \bigg) \bigg\}.
\end{aligned}
\end{equation} 

\subsection{$\mb$-minimization}
The sub-problem associated with optimization of $\mb$ is 
\begin{equation}
\begin{aligned}
& \min_{\mb}   \frac{\rho}{2} \| \mc - \textbf{D}\mb + \frac{1}{\rho}{\mz}^{\mc} \|_F^2  + \frac{\rho}{2} \| \ma - \mb + \frac{1}{\rho}{\mz}^{\mb} \|_F^2	\\
\end{aligned}
\end{equation} 
which can be solved with the following equation:
\begin{equation}
\begin{aligned}
(\textbf{D}^T\textbf{D}+I)\mb = \textbf{D}^T (\mc+\mz^{\mc}/\rho) + (\ma + \mz^{\mb} / \rho),
\end{aligned}
\end{equation} 
which can be efficiently solved by the fast Fourier transform (FFT):
\begin{equation}
\label{eq_sol_B}
\mb = \mathcal{F}^{-1} \left[ \frac{\mathcal{F} (\textbf{D}^T (\mc+\mz^{\mc}/\rho) + (\ma + \mz^{\mb} / \rho) ) }
{ 1+ \mathcal{F}(\tau_x\textbf{D}_x)^2 + \mathcal{F}(\tau_y\textbf{D}_y)^2 + \mathcal{F}(\tau_z\textbf{D}_z)^2 } \right].
\end{equation}


\subsection{$\mc$-minimization}
The sub-problem associated with optimization of $\mc$ is
\begin{equation}
\begin{aligned}
& \min_{\mc} \gamma \|\mc\|_{1} + \frac{\rho}{2} \| \mc - \textbf{D}\mb + \frac{1}{\rho}{\mz}^{\mc} \|_F^2	\\
\end{aligned}
\end{equation} 
The above problem can be solved with the soft-shrinkage operator in an element-wise manner, which leads to 
\begin{equation}
\label{eq_sol_C}
\mc_{i,j} = \max \big( (\textbf{D}\mb - {\mz}^{\mc} / \rho)_{i,j}  - \gamma / \rho, 0 \big).
\end{equation}

\subsection{Updating of $\mz^{\mo}_{i,j}$, $\mz^{\ma}_{i,j}$, $\mz^{\mb}$, $\mz^{\mc}$, and $\rho$}
We update the following variables in a standard way:
\begin{equation}
\label{eq_sol_lag}
\begin{aligned}
\mz^{\mo}_{i,j} = & \mz^{\mo}_{i,j} + \rho(\mo_{i,j} - \ml_{i,j} - \ms_{i,j}),	\\
\mz^{\ma}_{i,j} = & \mz^{\ma}_{i,j} + \rho(\ml_{i,j} - \ma_{i,j}),	\\
\mz^{\mb} = & \mz^{\mb} + \rho(\ma-\mb),	\\
\mz^{\mc} = & \mz^{\mc} + \rho(\mc-\textbf{D}\mb),	\\
\rho = & \rho\kappa, 
\end{aligned}
\end{equation}
where $\kappa > 1$ is a parameter that keeps $\rho$ increasing along with the optimization iterations.
We summarize the above optimization strategy in \ref{algorightm-1}.

{
To analyze the complexity of the proposed algorithm, we first consider the complexity of updating each patch at each iteration as follows. 
To update $\ml_{i,j}$ with \cref{eq_sol_L}, the major complexity comes from the computation of SVD, which has a complexity of $O(\textit{min}(mn^2,m^2n))$.
To update $\ms_{i,j}$ with \cref{eq_sol_S}, the complexity is $O(mn)$.
To update $\ma_{i,j}, \mb_{i,j}$, and $\mc_{i,j}$ with \cref{eq_sol_A,eq_sol_B,eq_sol_C}, the complexity for each variable is $O(mn)$, $O(mn\log(mn))$, and $O(mn)$, respectively.
To update $\mz^{\mo}_{i,j},\mz^{\ma}_{i,j}, \mz^{\mb}$, and $\mz^{\mc}$ by \cref{eq_sol_lag}, the complexity is $O(m,n)$. 
In summary, the overall complexity for each patch at each iteration is $O(\textit{min}(mn^2,m^2n))+mn\log(mn))$.
Thus, the overall complexity of the proposed algorithm is $O(\frac{MN}{mn}t_{max}(\textit{min}(mn^2,m^2n)+mn\log(mn))) = O(MN(\textit{min}(m,n)+\log(mn))t_{max})$.
Often times, for the path size we have $m=m$ and thus the overall complexity can be reduced to $O(MNnt_{max})$.
}

\begin{algorithm}
	\scriptsize
	\caption{HSI restoration via L$^3$S$^3$TV model.} \label{algorightm-1}
	\begin{algorithmic}[1]
		\REQUIRE
		Observed HSI $\mo\in\mathcal{R}^{M\times N\times p}$, patch size $m \times n$, 
		stopping criterion $\epsilon$, maximum number of iterations $t_{max}$, balancing parameters $\lambda$, $\gamma$, $\tau$, parameters $\rho_{max}$, $\rho^{(0)}$, $\kappa$.
		\STATE Initialize:  $\ml^{(0)},\ms^{(0)},\ma^{(0)},\mb^{(0)},(\mz^{\mo})^{(0)},(\mz^{\ma})^{(0)},(\mz^{\mb})^{(0)},\mc^{(0)},(\mz^{\mb})^{(0)}$, $t=1$.
		\STATE Repeat
		\STATE $\quad$ Update all patches $\big((\mathcal{L}_{i, j})^{(t)}, (\mathcal{S}_{i, j})^{(t)}\big)$ by \cref{eq_sol_L} and \cref{eq_sol_S}, respectively;
		\STATE $\quad$ Update $\ma^{(t)}, \mb^{(t)}, \mc^{(t)}$ by \cref{eq_sol_A,eq_sol_B,eq_sol_C}, respectively;
		\STATE $\quad$ Update $(\mz^{\mo})^{(t)},(\mz^{\ma})^{(t)},(\mz^{\mb})^{(t)}$, and $(\mz^{\mc})^{(t)}$ by \cref{eq_sol_lag};
		\STATE $\quad$ Update $\rho^{(t)} :=\min \left(\rho^{(t-1)} \kappa, \rho_{\max }\right)$
		\STATE $\quad$ Check the convergence condition:\\
			$\quad$ $\max \big\{\|\mo_{i, j} - \ml_{i, j}^{(t)}-\ms_{i, j}^{(t)}\|_{\infty}, \|\ml^{(t)}-\ma^{(t)}\|_{\infty},
			\|\ma^{(t)} - \mb^{(t)}\|_{\infty},$ \\
			$\quad$ $\quad$ $\quad$   $\|\mc^{(t)} - \textbf{D}\mb^{(t)}\|_{\infty} \big\} \!\leq \varepsilon$ or $t\ge t_{max}$.
		\STATE $\quad$ $t = t+1$. 
		\ENSURE
		Denoised image $\mathcal{L} = \ml^{(t)}$;
	\end{algorithmic}
\end{algorithm}

{
\section{Perturbation Analysis}
In the objective function given in \cref{eq_obj}, the log-determinant and $\ell_{2,\log}$ norm are nonconvex. 
In general, the algorithm of L$^3$S$^3$TV can only find a local minimum. Inspired by 
\cite{laurberg2008theorems}, we can formulate the estimation error between the unknown global minimum 
and the computationally obtained local minimum as estimation noise. This formulation will need  
the following lemma. 
\\

\begin{lemma}
The $\ell_{2,\log}$-norm satisfies triangle inequality.
\end{lemma} 
\noindent{{\em{Proof}}}:
Let $A$ and $B$ that be two matrices of the same size.
Denote the $j$-th column of matrix $X$ by $X_j$. Then, we have 
\begin{align*}
& \| A \|_{2, \log} + \| B \|_{2, \log} \\
= &  \sum\nolimits_j \log (1+ \| A_{j} \|_2 ) + \log (1+ \| B_{j} \|_2 ) \\
= &  \sum\nolimits_j \log (1+ \| A_{j} \|_2 + \| B_{j} \|_2 + \| A_{j} \|_2  \| B_{j} \|_2 ) \\
\ge &   \sum\nolimits_j \log (1+ \| A_{j} + B_{j} \|_2 ) = \| A  +  B \|_{2, \log}. \qquad\qquad\quad \Box 
\end{align*}

By using this property of the $\ell_{2,\log}$-norm, we can bound the estimation 
noise at a certain level as follows. 
Without loss of generality, we consider an arbitrary patch of \cref{eq_obj} and then the conclusion can be generalized to the overall objective.

\begin{theorem}
	Given a nontrivial observation data matrix $O$,
	let $L^*$ and $S^* = O - L^*$ be a pair of global solution minimizing the objective in $L$ 
	given in \cref{eq_obj}. 
	For any sufficiently small $\epsilon > 0$, there exists an $\delta>0$ such that for any non-negative observation 
	matrix $\bar{O} = O + E$ with $\| E \|_{2, \log} \le \delta$, we  have 
	\begin{equation}
	\nonumber J_{L^*} (L) :=  \| L^* - L \|_{\log\det} + \lambda \| L^* - L \|_{2, \log} + \mu \ 
	\| L^* - L  \|_{\sstv} \ < \epsilon,
	\end{equation} 
	where $L$ is a local solution of  $\bar{O}$ estimated by the algorithm for solving \cref{eq_obj}. 
\end{theorem}

\noindent{{\em{Proof}}}:
First, define the main objective function of \cref{eq_obj} to be 
$R(L; O)$. 
Without loss of generality, we will consider $\lambda = 1$ and other $\lambda$ values can be similarly treated.
Now  take a sufficiently small $\epsilon$ which is $0 < \epsilon < \| L^* - O \|_{2, \log}$, 
and let 
$\delta_1 = R(L^*; O)$. Note that $\delta_1 > 0$, because $\delta_1 = 0$ is equivalent to $O=0$, violating the condition on $O$. 
Necessarily we have $\epsilon < \delta_1$.

Let ${\mathcal{G}}$ be the open set of all $L$ close to $L^*$: 
\[
\nonumber {\mathcal{G}} := \{ L \ | \ J_{L^*} (L) < \epsilon \}.
\]
Also, define a set $B := \{ L \ | \ \max (L) \le \text{e}^{\ \xi + \delta_1} - 1 + \max( O ) \}$, 
where $\xi$ is a positive constant. Let $\bar{{\mathcal{G}}} := \{ L \ | \ J_{L^*} (L) \ge \epsilon \} \cap B$.
Because $\bar{{\mathcal{G}}}$ is bounded and closed and the norms are continuous, we have 
$\min_{L \in \bar{{\mathcal{G}}}} \{ R(L; O) - R(L^*; O)  \} = 
\delta^{\prime} > 0$.

For $\hat{L} \notin {\mathcal{G}}$ and $\max (\hat{L}) > \text{e}^{\ \xi + \delta_1} - 1 + \max( O )$, 
it is straightforward to verify that  
$\| O - \hat{L} \|_{2, \log} = \sum_{j} \log(1+ (\sum_i (O_{i,j} - \hat{L}_{i,j})^2)^{1/2}) \ge \log(1+\max(\hat{L}) - \max(O)) > \xi + \delta_1$. 
Here $\max(A)$ operates on the set of all elements of a matrix $A$. 
Therefore, $R(\hat{L}; O) - R(L^*; O) \ge \|\hat{L} - O\|_{2,\log} - R(L^*; O) > \xi +\delta_1 - \delta_1 = \xi$. 
Now take ${\delta} = \min \{ \xi, \delta^{\prime} \} / 3 $, 
then we have $R(\hat{L}; O) - R(L^*; O) \ge 3 {\delta}$, 
for any $\hat{L} \notin {\mathcal{G}}$.

For any $\bar{O}$ such that $\| \bar{O} - O \|_{2, \log} < \delta$, by the triangle  inequality of $(2, \log)$ norm, we have 
$R(\hat{L}; \bar{O}) + \| O - \bar{O} \|_{2, \log} 
\ge R(\hat{L}; O)$. Similarly, we have 
$R(L^*; \bar{O}) + \| O - \bar{O} \|_{2, \log} \ge R(L^*; O)$. 
Therefore, we have 
\begin{align}
\nonumber R(\hat{L}; \bar{O}) - R(L^*; \bar{O}) 
&\ge \  R(\hat{L}; O) - \| O - \bar{O} \|_{2, \log} - R(L^*; \bar{O}) \\
\nonumber &\ge \  R(\hat{L}; O) - R(L^*; O) - 2 \| O - \bar{O} \|_{2, \log} \\
& > 3 {\delta} - 2 \delta = \delta. 
\end{align}
All solutions that are not in ${\mathcal{G}}$ will not be a minimizer of the objective function. This concludes the proof.
$\hfill \Box$

This theorem asserts that the discrepancy between $(L^*, S^*)$ and the 
$(L, S)$ exists when the observation has additive noise; nonetheless, 
the estimation error will be small if the noise level is below a certain level. 
In this sense, we may regard the local solution of \cref{eq_obj} found by our algorithm
as a global solution with a small level of additive noise. 

{\bf{Corollary}} {\sl{
		Let $L^*$ and $L$ respectively be the global and local solutions obtained from the optimization problem defined in \cref{eq_obj}, where $L = L^* + E$, with $E$ being some errors. 
		For a small positive $\epsilon$, there exists a positive $\delta$ such that if $\max{E}$ is bounded by $\delta$, then we have that the discrepancy $J_{L^*}(L) < \epsilon$.  
	}}
}

\begin{table*}[t]
	\caption{Quantitative evaluation of different methods in different noise cases of Indian Pines dataset}
	\centering \scriptsize
	\label{tab_comp}
	\begin{tabular}{ccrrrrrrrrrr}
		\hline
		Noise  & Metric  & \quad BM4D  & \quad LRMR &  NAILRMA    & \quad U-FFP  & \qquad RPCA 
		& \quad LRTV & \quad LRTDTV  & \quad\quad SSTV  & \quad LLRGTV & \quad Ours \\
		\hline 
		\multirow{4}{*}{Case 1} 
		& MPSNR  & 37.985  & 37.032 & 37.234 & 35.722 		& 32.279 	& 39.316 &\underline{40.569} & 32.857 &40.497 & \bf{42.809}\\
		& MSSIM  & 0.970   & 0.946  & 0.944  & 0.917  		& 0.880  	& 0.986 &\underline{0.989} & 0.853  &0.976  &\bf{0.993} \\
		& ERGAS  & 30.490  & 33.902 & 32.872 & 39.174 		& 58.783 	& 26.722 &23.958 & 57.775 &\underline{22.860} &\bf{17.417} \\
		& TIME(s)& 194.973 & 209.638& 167.365&\bf{66.960} 	& 1204.579 	&\underline{117.384} &129.790 &171.802 &148.514 &134.074 \\
		\hline
		\multirow{4}{*}{Case 2} 
		& MPSNR  & 34.530  & 34.251 & 36.038 & 35.351 & 32.018 &39.034 &\underline{39.996} & 32.003& 38.154 & \bf{40.303} \\
		& MSSIM  & 0.933   & 0.920  & 0.937  & 0.916  & 0.878  &0.987 &\underline{0.989} & 0.842 &0.973   &\bf{0.990} \\
		& ERGAS  & 112.266 & 62.896 & 40.614 & 40.797 & 60.296 &\underline{27.254} &\bf{25.230} & 63.943 &45.522  &31.870 \\
		& TIME(s)&195.659 &210.100 &154.193 &\bf{71.151} &2137.155 &91.433 &390.807 &180.881 &\underline{77.803} &130.683 \\
		\hline
		\multirow{4}{*}{Case 3} 
		& MPSNR  & 34.608 & 33.825 & 34.101 & 32.599 & 29.710 &37.491 &\underline{38.037} & 30.431 & 37.707 & \bf{39.184} \\
		& MSSIM  & 0.927  & 0.902  & 0.899  & 0.855  & 0.810  &0.980 &\bf{0.983} & 0.774  &0.959  &\underline{0.981} \\
		& ERGAS  & 51.032 & 49.898 & 48.082 & 56.191 & 78.723 &33.362 &31.923 & 74.370 &\underline{31.978}  &\bf{27.355} \\
		& TIME(s)&195.991 &210.951 &171.461 &\bf{62.334} &1650.750 &\underline{79.891} &270.448 &220.813 &151.789 &132.173 \\
		\hline
		\multirow{4}{*}{Case 4} 
		& MPSNR  & 34.670 & 35.328 & 37.622 & 37.215 & 33.973 &41.427 &\bf{41.871} & 33.471 & 39.241 & \underline{41.629} \\
		& MSSIM  & 0.909  & 0.933  & 0.950  & 0.945  & 0.904  &0.991 &\underline{0.993} & 0.883  &0.975   &\bf{0.993} \\
		& ERGAS  & 111.642& 59.989 & 35.388 & 34.075 & 48.583 &\underline{21.400} &\bf{20.886} & 57.185 &33.039  &30.565 \\
		& TIME(s)&196.357 &211.355 &157.418 &\bf{66.837} &654.329 &98.077 &248.448 &217.216 &\underline{80.850} &128.464 \\		
		\hline
		\multirow{4}{*}{Case 5} 
		& MPSNR  & 36.210 & 37.136 & 37.043 & 37.628 & 34.254 &41.705 &\underline{42.067} & 33.454 & 42.473 & \bf{42.906} \\
		& MSSIM  & 0.947  & 0.949  & 0.959  & 0.946  & 0.910  &0.992 &\underline{0.993} & 0.872  &0.986   &\bf{0.993} \\
		& ERGAS  & 37.186 & 33.530 & 26.622 & 31.393 & 46.680 &20.303 &20.229 & 54.544 &\underline{20.209}  &\bf{17.270} \\
		& TIME(s)&225.938 &211.394 &176.956 &205.189 &915.991 &\bf{72.276} &237.115 &141.643 &\underline{82.000} &112.715 \\
		\hline
		\multirow{4}{*}{Case 6} 
		& MPSNR  & 32.617 & 30.804 & 31.362 & 30.073 & 27.376 &34.306 &\underline{35.514} & 27.998 & 35.322 & \bf{35.750} \\
		& MSSIM  & 0.912  & 0.846  & 0.845  & 0.771  & 0.730  &\bf{0.970}  &\underline{0.967}  & 0.678  &0.931  &0.959 \\
		& ERGAS  & 56.346 & 69.820 & 69.820 & 75.114 & 102.106 &47.091 &41.817 & 96.043 &\underline{41.748}  &\bf{39.311}\\
		& TIME(s)&194.592 &202.756 &163.278 &130.246 &1103.551 &\bf{74.579} &193.667 &208.027 &\underline{84.442} &129.099 \\
		\hline 
	\end{tabular}
	\\ The best performance is \textbf{boldfaced} while the top second one is \underline{underlined}.  
\end{table*}

\section{Experiments}
\label{sec_exp}
In this section, we conduct extensive experiments to testify the effectiveness of the proposed method.
In particular, we compare our method with seven state-of-the-art HSI denoising methods, 
including the block-matching and four-dimensional filtering algorithm (BM4D) \cite{maggioni2013nonlocal}, 
the convex approach to RPCA \cite{Cand2011Robust}, 
GoDec-based low-rank matrix recovery (LRMR) \cite{zhang2014hyper}, 
spatial-spectral total variation (SSTV) \cite{chan2011an}, 
noise-adjusted iterative low-rank matrix approximation (NAILRMA) \cite{he2015Hyperspectral}, 
factorization-based nonconvex RPCA (U-FFP) \cite{peng2020robust}, total variation regularized low-rank matrix factorization (LRTV) \cite{he2016total}, and local low-rank matrix recovery with global spatial-spectral total variation (LLRGTV) \cite{he2018hyperspectral}. 
We follow a strategy similar to the literature \cite{he2018hyperspectral,he2015Hyperspectral} and scale the gray value of each band of HSIs to the interval [0,1] before denoising.
After denoising, we restore the recovered images to the original gray value level, 
which facilitates numerical calculation and visualization. 
In the experiments, we tune the parameters for each method such that they are manually adjusted to the best according to the default strategy. 

\subsection{Simulated HSI Data Experiments}
\label{sec_data_syn}

In this test, we conduct experiments on simulated data sets to quantitatively compare all methods in HSI denoising.
In particular, we use the ground truth image of the Indian pines data set to generate the synthetic data, which has a size of 145$\times$145$\times$224. 
We treat the synthetic data set as the ground truth (GT) of the simulated HSIs.
Then, we use the GT of the simulated HSIs to generate noisy HSIs, where we artificially add noise to the GT under 6 conditions, resulting in 6 noisy HSI data sets.
We describe how we add noise to the GT to obtain the noisy data sets as follows:

\begin{figure}[t]
	\centering
	{\includegraphics[width=0.95\columnwidth]{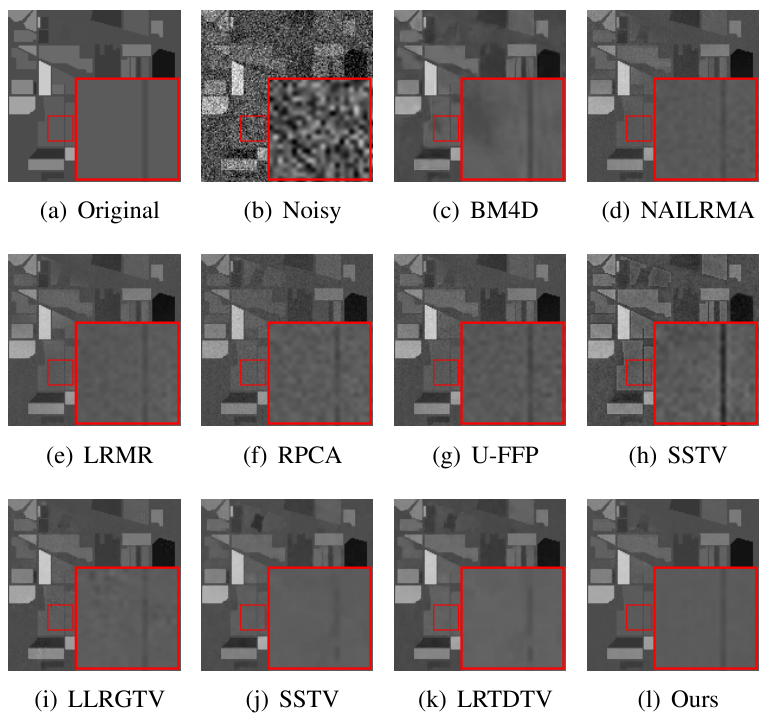} }	\\ 
	\caption{ Restoration results on synthetic data set of Case 3. (a) is the 206th band of GT; (b)-(k) are the restored results of the 206th band by different methods. }
	\label{fig_syn_200}
\end{figure}

\begin{itemize}
\item Case 1: We add zero-mean Gaussian noise to each band of the GT, where all bands are fixed to have the same noise intensity. In this case, we set the noise variance to be 0.1;

\item Case 2: Based on Case 1, we further add some deadlines to bands 81-120. In each of these bands, we randomly add 3-10 deadlines, where each deadline has a random width of 1-3 columns;  

\item Case 3: Similarly to Case 1, we add Gaussian noise to the GT, except that the noise variance is set to 0.14 in this case. Then we add some stripes in bands 161-190. 
In each of these bands, we randomly select 20-40 columns to add stripes.
For each column, we randomly select a value within [-0.25, 0.25] and add it to all pixels of this column;

\item Case 4: Based on Case 2, we further add some stripes in bands 161-190 in the same way as in Case 3;

\item Case 5: First, we randomly add zero-mean Gaussian noise with an SNR value between 15 and 25 dB to each band separately, where the average SNR value of the noise in all bands is 20.43 dB. 
Then we add salt and pepper impulse noise to the data set by randomly corrupting 20\% pixels in each band. 
Finally, we randomly add impulse noise with intensity between 0.0196 and 0.0784 to each band separately, where the average intensity of the impulse noise in all bands is 0.0492;

\item Case 6: First, we randomly add zero-mean Gaussian noise with an SNR value between 45 and 55 dB to each band separately, where the average SNR value of the noise in all bands is 49.75 dB. 
Then we add impulse noise to the data set in the same way as in Case 5. 

\end{itemize}
In the rest of this subsection, we will conduct extensive experiments to test the proposed method both quantitatively and qualitatively. The detailed comparison results and discussions are presented in the following.

\begin{figure*}[!t]
	\centering
	{\includegraphics[width=1.9\columnwidth,height=0.55\columnwidth]{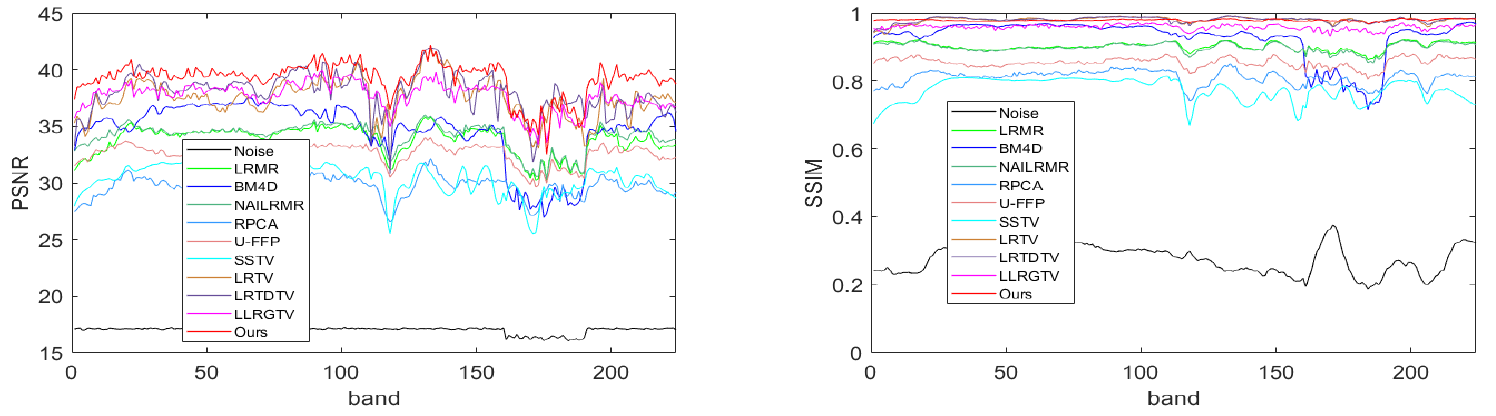} }
	\caption{ Comparison of different methods in PSNR (on left) and SSIM (on right) on each band of Case 3, respectively. }
	\label{fig_bands_test}
\end{figure*}

\begin{figure}[!t]
	\centering
	\includegraphics[width = 0.95\columnwidth]{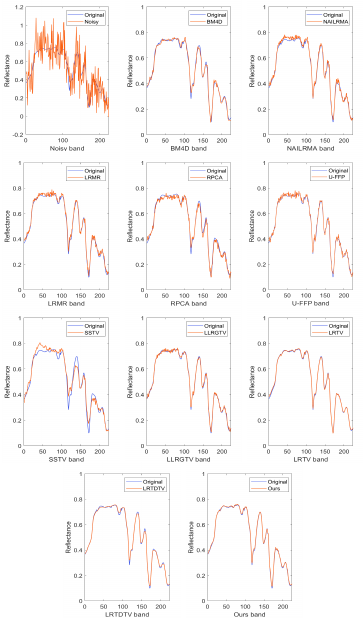}
	\caption{Spectrum of pixel (140, 90) in the restoration results of synthetic data  in Case 3.}
	\label{fig_spectrum_140_90}
\end{figure}

To quantitatively evaluate the proposed method, we adopt three widely used evaluation metrics to compare all methods, 
including the peak signal-to-noise ratio (PSNR), structural similarity (SSIM), and relative global scale (ERGAS).
Better performance is obtained with higher values for the first two metrics while lower values for the last one, respectively. 
Since the first two metrics are applied to every single band, we report the averaged values over all bands, which are denoted as mean PSNR (MPSNR) and mean SSIM (MSSIM), respectively.
We apply all methods to the above generated noisy data sets and report the detailed denoising performance in \cref{tab_comp}.
{
It is seen that the proposed method and LRTDTV are among the top-tier methods, where they obtain almost all the top two denoising performances.
In particular, the proposed method achieves the top one and two performances in 13 and 15 out of 18 cases, respectively,
which shows its superiority to other baseline methods.
In other cases, the proposed method also has quite competitive performances.
Compared with LRTDTV, the proposed method has significant improvements in MPSNR and ERGAS.
For example, the proposed method improves the performance by about 2 and 6 in MPSNR and ERGAS in Cases 1, respectively.}
Compared with the other baseline methods, the improvements of the proposed method are more significant.
For example, BM4D, LRMR, and NAILRMA are among the second-tier methods. 
Compared with them, the proposed method has improvements by about 5 in MPSNR in almost all cases. 
In ERGAS, the proposed method improves the performance by at least 10 in almost all cases, and the improvement can be even about 80 on Case 2. 
In MSSIM, the proposed method improves the performance by about 0.05 in almost all cases. 
{
Regarding time cost, the proposed method is quite competitive. Several methods obtain the results with least time cost in different cases. 
However, none of them costs the least time in all cases.
Generally, among the methods with good performance, the proposed method has quite competitive speed.
Although methods such as U-FFP may have faster speed, their denoising performances are far worse than the proposed method.
Considering the superior denoising performance, such speed is indeed acceptable for the proposed method.}
These observations suggest the effectiveness and superior performance of the proposed method to the baseline methods from a quantitative perspective.

To evaluate the proposed method from a qualitative perspective, we further show some visual results for detailed comparison.
For all methods in comparison, without loss of generality, we show the denoised 206th band of Case 3 in \cref{fig_syn_200}.
It is seen that among all methods, the NAILRMA, LRMR, RPCA, U-FFP, SSTV, and LLRGTV fail to remove the mixed types of noise. 
BM4D has relatively better performance with the majority noise removed; 
however, it has over-smoothing effects in the recovered image, where the edges between smooth regions are blurred, and some adjacent smooth regions are merged. 
{LRTV and LRTDTV are among the top methods, where it is seen that they eliminate the majority noise and well captures the detailed information of regions.
However, the edge information is damaged in these images and shows inferior performance compared with the proposed method.
Among all methods, the proposed method obtains the best denoising performance, where it is seen that the restored image contains smooth regions, clean edge structures, and eliminates noise.}
These observations confirm the effectiveness of the proposed method from a visual perspective.

In \cref{tab_comp}, we have compared all methods with the averaged performance over all bands in MPSNR, MSSIM, and ERGAS, respectively.
To better compare the methods in our experiment, in this test, we further compare their performances on each individual band. 
Without loss of generality, we show the results of Case 3 in terms of PSNR and SSIM in \cref{fig_bands_test}, respectively. 
{It is noted that the proposed method achieves the top performance in PSNR with significant improvements in almost all bands.
In SSIM, the proposed method is quite competitive among all methods and has at least the top second performance in almost all bands.}
This suggests that the proposed method not only achieves promising performance on the overall data set but also in each band, which is essentially important in real-world applications.

To further assess the effectiveness of the proposed method, in this test, we show some results of the spectral signatures in the restored images.
In particular, we show the spectral signature curves of the restored images in Case 3 at the location of pixel (140, 90) across all bands.
To better investigate the quality of the restored images by each method, we show the spectral signature curves of both the restored images and the GT in \cref{fig_spectrum_140_90}.
For the recovered images with high quality, their spectral signature curves are expected to be very close to the those of the GT.
{
Among the baseline methods, the LLRGTV, LRTV, and LRTDTV have the best performances, where we observe that their curves are the closest ones to the original curve.
This confirms the relatively superior performance of these methods among all baseline methods.
Compared with these methods, the proposed method has even better performance.
Specifically, the curve obtained by LLRGTV is more zigzag than that by our method, especially in bands 20-100, implying that the proposed method better retains spectral consistency between nearby bands in the recovered HSI.
The curve obtained by LRTDTV does not has zigzag effects, but it fits the original curve less closely compared with the curve obtained by our method.
LRTV has relatively better performance than LRTDTV and LLRGTV, but inferior performance to the proposed method, especially in the first 100 bands. }
For the other methods, the curves of the recovered HSIs have significant differences from the original HSI, which show obvious inferior performance to the proposed method.
These observations confirm the effectiveness and superior performance of the proposed method.

\begin{figure}[t]
	\centering
	{\includegraphics[width=0.9\columnwidth]{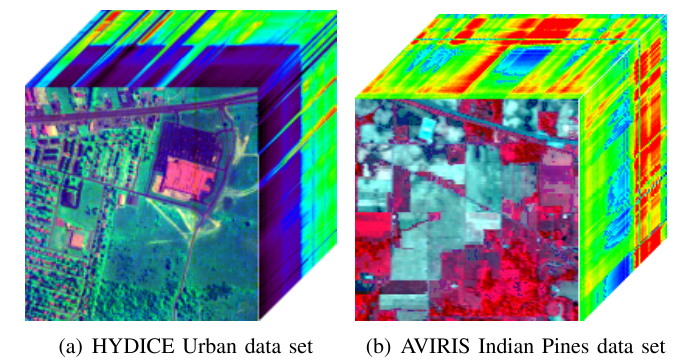} }
	\caption{ Examples of real-world data sets used in the experiment: (a) HYDICE Urban data set  (R: 6, G: 88, B: 210); (b) AVIRIS Indian Pines (R: 60, G: 27, B: 17). }
	\label{fig_realdata}
\end{figure}

\subsection{Experiments on Real World HSI Data Sets}

{
In this test, we conduct experiments on three real-world HSI data sets, 
including the hyperspectral digital image acquisition experiment (HYDICE) urban data set, the airborne visible infrared imaging spectrometer (AVIRIS) Indian pines data set, and the EO-1 Hyperion Australia data set (Hyperion).
For a visual illustration, we show some examples of the false-color HSI in \cref{fig_realdata}. }
In the rest of this subsection, we will present brief descriptions of the data sets and the detailed experimental results, respectively.

\begin{figure}[t]
	\centering
	{\includegraphics[width=0.95\columnwidth]{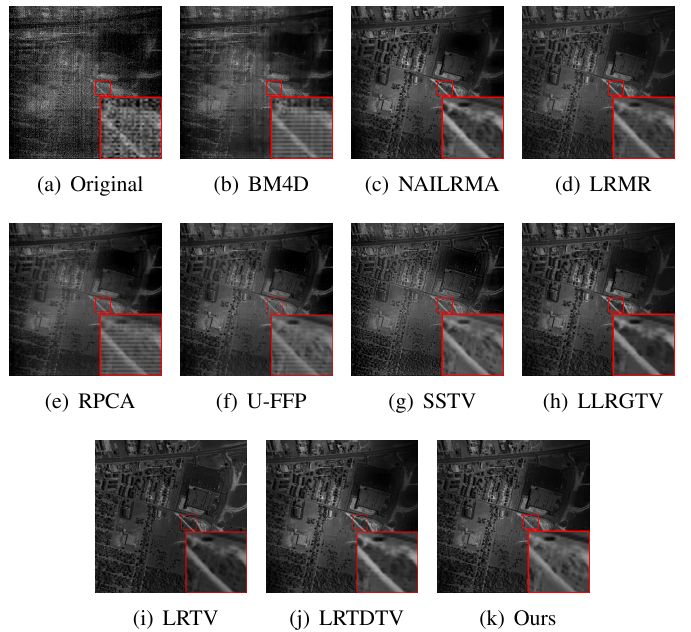} }
	\caption{ Restoration results on HYDICE urban data set. (a) is the original 108th band of HYDICE; (b)-(k) are the restored results of the 108th band by different methods. The figure is better viewed in a zoomed-in PDF.}
	\label{fig_urban108}
\end{figure}

\begin{figure}[t]
	\centering
	{\includegraphics[width=0.95\columnwidth]{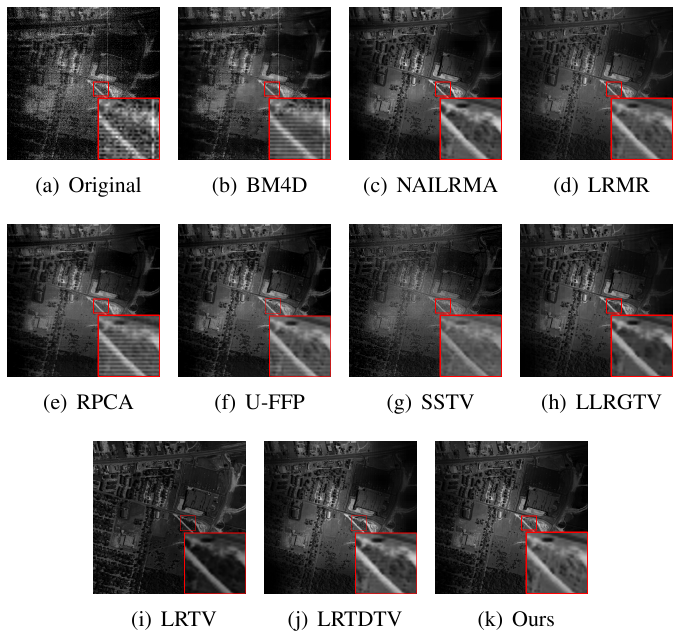} }
	\caption{ Restoration results on HYDICE urban data set. (a) is the original 139th band of HYDICE; (b)-(k) are the restored results of the 139th band by different methods. The figure is better viewed in a zoomed-in PDF.}
	\label{fig_urban139}
\end{figure}

\subsubsection{HYDICE Urban Data Set}
\label{sec_exp_urban}

The HYDICE data set contains 210 bands of images with a size of 307$\times$307 pixels. 
This data set is heavily corrupted with stripes, deadlines, atmosphere, water absorption, and other unknown types of noises.
In our experiment, we test the methods using all bands of the data set.

Since the real-world data sets lack clean bands, it is not straightforward to evaluate the methods quantitatively.
Thus, all methods are evaluated from a visual quality perspective, which suggests us to tune the parameters for the methods as follows.
For RPCA, we tune its parameter around the theoretically optimal value to obtain the best visual performance.
For the other methods, we follow a common strategy and manually tune their parameters such that the best visual performance is observed for each method. 

Without loss of generality, we show the results of the 108th and 139th bands obtained by each method in \cref{fig_urban108,fig_urban139}, respectively.
Compared with other methods, BM4D and RPCA have inferior performance on this data set,
where it is observed that both the 108th and 139th bands restored by BM4D and RPCA have strong fringe noise effects. 
The restored images by the NAILRMA, LRMR, U-FFP, { and LRTDTV} also have fringe effects in both bands, but they are much lighter than those restored by BM4D and RPCA. 
SSTV removes almost all noise in band 108, but we can still observe fringe effects and some detail information is missing in band 139. 
{Among all methods, LLRGTV, LRTV and our method are the only ones that well remove the fringes from the HSIs.
It is seen that LLRGTV and LRTV generate clearly visible images and appear effective in removing noise. 
Unfortunately, both LLRGTV and LRTV fail to retain rich detail information from the noisy HSIs as our method does.}
For example, in the amplified regions, it is seen that the region is over-smoothed and structural detail information is missing.
{
Moreover, the images generated by LRTV appears darker than the other images as well as the original one.}
In summary, on the HYDICE data set, the proposed method removes the noise and well retains local details while the baseline methods fail, 
suggesting superior performance and effectiveness of the proposed method in HSI denoising.

\begin{figure}[t]
	\centering
	{\includegraphics[width=0.95\columnwidth]{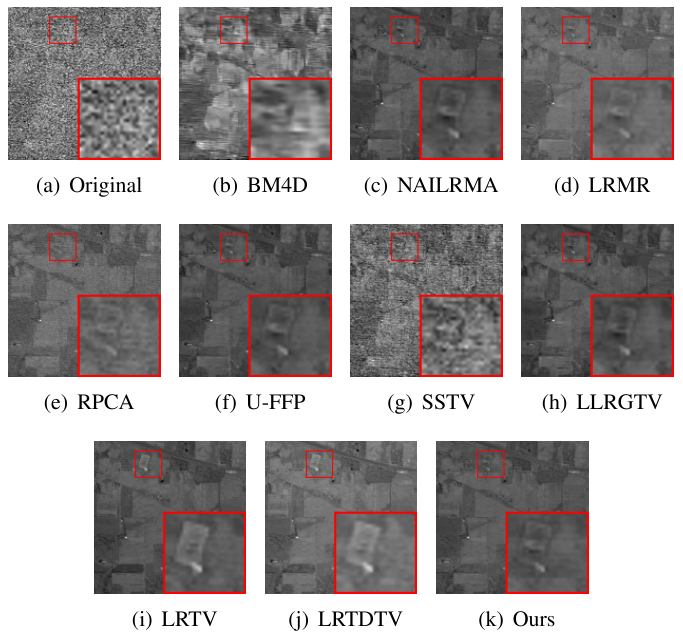} }
	\caption{ Restoration results on AVIRIS Indian Pines data set. (a) is the original 150th band of AVIRIS; (b)-(i) are the restored results of the 150th band by different methods. The figure is better viewed in a zoomed-in PDF.}
	\label{fig_avirs150}
\end{figure}

\begin{figure}[h]
	\centering
	{\includegraphics[width=0.95\columnwidth]{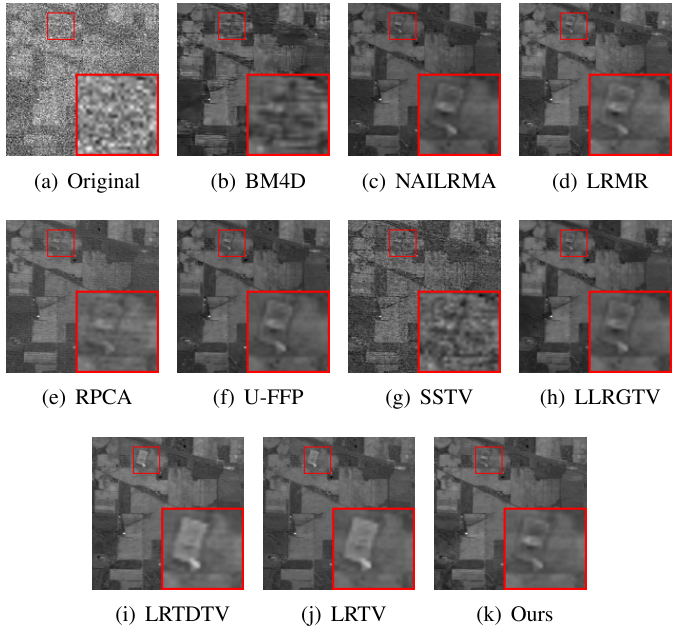} }
	\caption{ Restoration results on AVIRIS Indian Pines data set. (a) is the original 220th band of AVIRIS; (b)-(i) are the restored results of the 220th band by different methods. The figure is better viewed in a zoomed-in PDF.}
	\label{fig_avirs220}
\end{figure}

\subsubsection{AVIRIS Indian Pines Data Set}

The AVIRIS Indian pines data set consists of 220 bands of images with a size of 145$\times$145 pixels. 
In this data set, some bands are severely damaged by mixed Gaussian and impulse noise, which makes it challenging to remove noise from this data set.
Among all the bands, we select two typical ones, including bands 150 and 220, to show the denoising performance of all methods. 
The restored images of these two bands are shown in \cref{fig_avirs150,fig_avirs220}, respectively.
All parameters are tuned in a way that follows \cref{sec_exp_urban} on this data set.

It is seen that BM4D, RPCA, and SSTV fail to remove the noise in both bands 150 and 220.
In particular, we can observe heavy noise in the restored bands by SSTV and heavy fringes effects in those restored by BM4D and RPCA.
Compared with them, the LRMR, U-FFP, NAILRMA and LRTDTV have relatively better performance, where they remove majority of noise.
However, we can still observe some fringes effects remaining in the restored bands. 
Moreover, compared with the proposed method, they lose some rich structural details.
Among all baseline methods, the LLRGTV, LRTV, and LRTDTV have relatively the best performances.
{
It is observed that these methods remove almost all noise from the bands.
However, their performances are still inferior to the proposed method.
For example, some regions have less smooth effects and some detail information is missing in the image recovered by LLRGTV;
it is observed that the image recovered by the proposed method has less noise and clearer edge information than those recovered by LRTV and LRTDTV.}
These observations confirm the effectiveness of the proposed method on this data set.

\begin{figure}[t]
	\centering
	{\includegraphics[width=0.95\columnwidth]{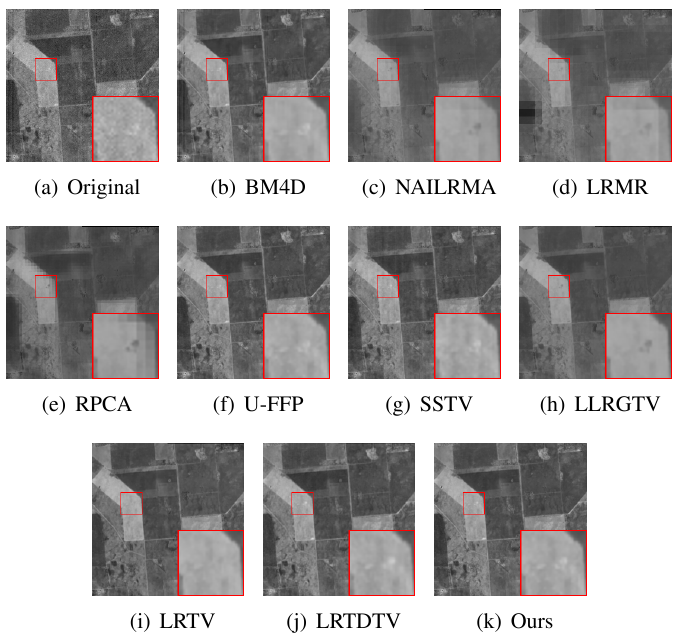} }
	\caption{ Restoration results on Hyperion data set. (a) is the original 47th band of Hyperion; (b)-(k) are the restored results of the 47th band by different methods. The figure is better viewed in a zoomed-in PDF.}
	\label{fig_aus47}
\end{figure}

\begin{figure}[!]
	\centering
	{\includegraphics[width=0.95\columnwidth]{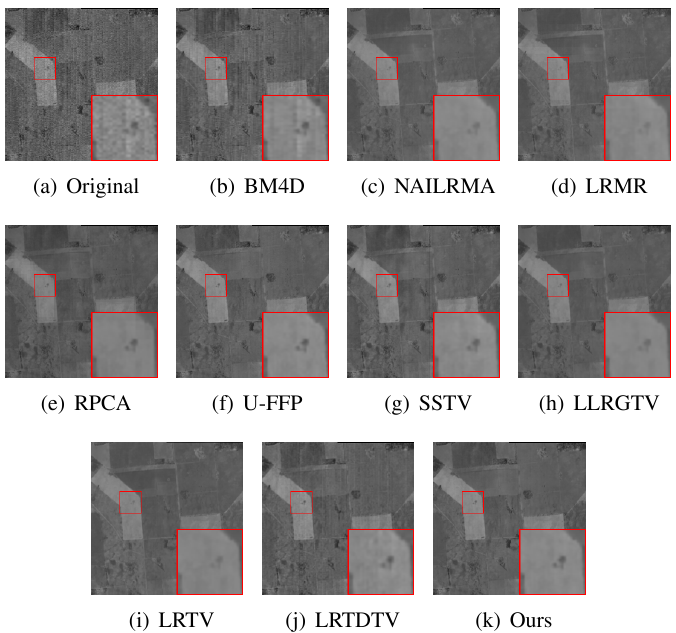} }
	\caption{ Restoration results on Hyperion data set. (a) is the original 89th band of Hyperion; (b)-(k) are the restored results of the 89th band by different methods. The figure is better viewed in a zoomed-in PDF.}
	\label{fig_aus89}
\end{figure}

{
\subsection{EO-1 Hyperion Australia Data Set}
The Hyperion image was captured on December 4, 2010, with original size of 3858$\times$256$\times$242.
We follow the strategy in the literature and remove the overlapping bands between visual near-infrared and shortwave infrared ranges.
As a result, we use a subregion of size 200$\times$200$\times$150 in our experiment.
Among all the bands, we select two typical ones, including bands 47 and 89, to show the denoising performance of all methods. 
The restored images of these two bands are shown in \cref{fig_aus47,fig_aus89}, respectively.
All parameters are tuned in a way that follows \cref{sec_exp_urban} on this data set.
In \cref{fig_aus47}, it is seen that images obtained by BM4D and SSTV still have some noise effects, 
while those obtained by NAILRMA, LRMR, RPCA, and LLRGTV have strong block or over smoothing effects.
Among all baseline methods, it is seen that LRTV, U-FFP, and LRTDTV have relatively better performance.
In the amplified region, it is seen that LRTV loses the detail information whereas the proposed method recovers the detail information better than U-FFP and LRTDTV.
For example, the proposed method recovers the detail information with clearer edges than U-FFP and LRTDTV.
In \cref{fig_aus89}, it is seen that BM4D, SSTV, and LRTDTV still have noise effects in the recoverd images while LRMR and RPCA have block effects. 
In the images recovered by NAILRMA, LLRGTV, and LRTV, the detail information is less clean than that recovered by the proposed method. 
The U-FFP has relatively the best performance among all baseline methods, but it has relatively more noise than the proposed method in the smooth regions of the recovered image.
These observations confirm the effectiveness of the proposed method on this data set.

{In summary, it is seen that none of the baseline methods shows the top performances on all these real-world data sets while the proposed method does, which confirms the effectiveness of the proposed method in HSI denoising.}

}


\begin{figure}[htbp]
	\centering
	{\includegraphics[width=0.8\columnwidth]{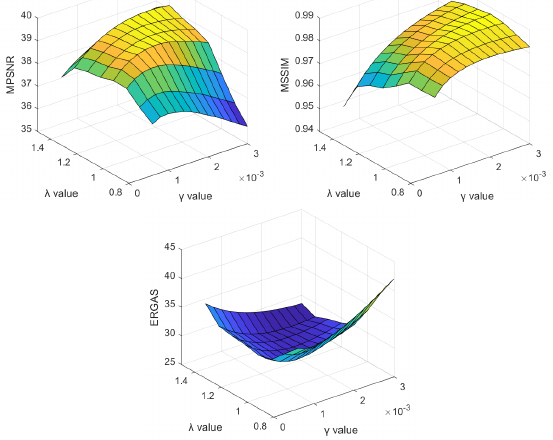} }
	\caption{The performance of the proposed method changes with respect to the parameter values in MPSNR, MSSIM and ERGAS, respectively. }
	\label{fig_parameter}
\end{figure}

\subsection{Parameter Sensitivity}
In the above tests, we have confirmed the effectiveness of the proposed method in HSI denoising.
In this test, we further show how the parameters affect the denoising performance of the proposed method. 
For a quantitative illustration and without loss of generality, we use the synthetic data in Case 3.
In particular, we report the performance of the proposed method in three metrics with respect to different combinations of $\lambda$ and $\gamma$ in \cref{fig_parameter}.
It is seen that the proposed method achieves relatively good performance when $\lambda$ is nearby 1 and $\gamma$ is small.
{
For such $\lambda$ and $\gamma$ values, it is seen that the proposed method achieves good performance with a wide range of combinations of parameters.
Similar observations can be found in other cases.
Such behavior is indeed important for unsupervised learning method to be potentially applied in real-world applications.
These observations suggest we adopt such values in real-world applications.
}

\begin{figure}[t]
	\centering
	{\includegraphics[width=0.95\columnwidth]{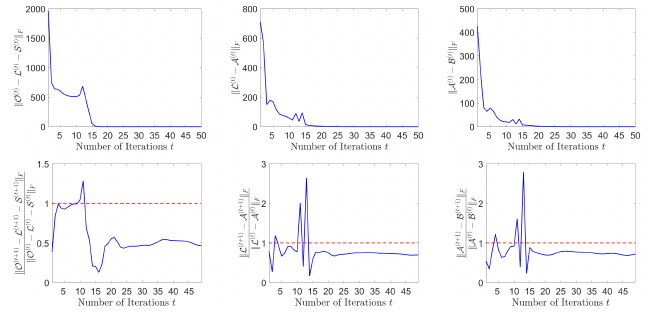} }
	
	\caption{ The convergence curves (on top) and the relative difference (on bottom) of consecutive updates generated by the proposed method on synthetic data in Case 3.  }
	\label{fig_conv_vars_relative_synthetic}
\end{figure}

\begin{figure}[!]
	\centering
	{\includegraphics[width=0.95\columnwidth]{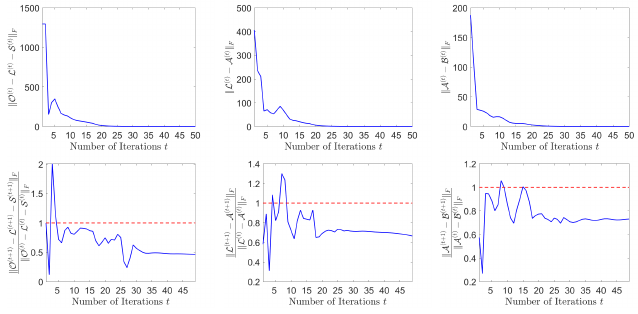} }
	\caption{The convergence curves (on top) and the relative difference (on bottom) of consecutive updates generated by the proposed method on Indian Pines data set.  }
	\label{fig_conv_vars_relative_Indian}
\end{figure}

\subsection{Convergence Study}
For the comprehensive optimization strategy, it is generally not straightforward to provide theoretical results on the convergence.
Thus, in this test, we empirically testify the convergence of the proposed method.
Without loss of generality, we first conduct experiments using the synthetic data Case 3, where the parameters are set to be $\lambda=1.3$ and $\gamma=0.0022$. 
For other values, generally we can observe similar patterns.

First, we show the convergence behavior of the proposed method in variable sequence by showing three different errors that measure how the constraints are met. 
{
Without loss of generality, we show how the updates of variables gradually satisfy the constraints by plotting the sequences of errors, 
including $\big\{\big\|\mo^{(t)}-\ml^{(t)}-\ms^{(t)}\big\|_F\big\}_{t=1}^{\infty}$, $\big\{\big\|\ml^{(t)}-\ma^{(t)}\big\|_F\big\}_{t=1}^{\infty}$, and $\big\{\big\|\ma^{(t)}-\mb^{(t)}\big\|_F\big\}_{t=1}^{\infty}$ in \cref{fig_conv_vars_relative_synthetic}, respectively. 
It is seen that the sequences of errors converge within a few numbers of iterations, indicating that the variable sequences gradually meet the constraints and converge. 
Moreover, to better show the convergence behavior, we also plot the sequences 
$\Big\{\frac{\|\mo^{(t+1)}-\ml^{(t+1)}-\ms^{(t+1)}\|_F}{\|\mo^{(t)}-\ml^{(t)}-\ms^{(t)}\|_F}\Big\}_{t=1}^{\infty}$, 
$\Big\{\frac{\|\ml^{(t+1)}-\ma^{(t+1)}\|_F}{\|\ml^{(t)}-\ma^{(t)}\|_F}\Big\}_{t=1}^{\infty}$, and
$\Big\{\frac{\|\ma^{(t+1)}-\mb^{(t+1)}\|_F}{\|\ma^{(t)}-\mb^{(t)}\|_F}\Big\}_{t=1}^{\infty}$ in \cref{fig_conv_vars_relative_synthetic}, respectively,
which show the relative changes between consecutive elements of sequences of errors.
It is seen that the relative changes are smaller than 1 after the first a few iterations.
To further show such behaviors on real-world data sets, we show some curves on the Indian Pines data set in \cref{fig_conv_vars_relative_Indian}, where we set $\lambda=1$ and $\gamma=0.0001$. 
It is seen that similar pattern to the synthetic data set is observed on real-world data set, 
which indicates that the sequences of errors indeed converge with at least a linear convergence rate, which is satisfactory and suggested efficiency in real-world applications.
}

\begin{figure}[htbp]
	\centering
	{\includegraphics[width=0.95\columnwidth]{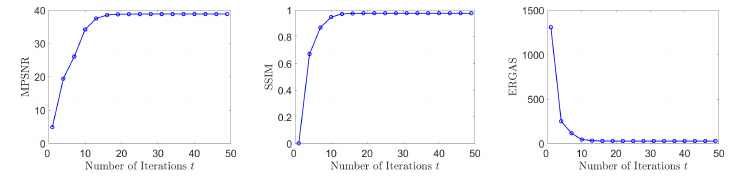} }
	\caption{The quality of denoised HSI in MPSNR, MSSIM, and GRGAS, respectively at different iterations in Case 3. }
	\label{fig_conv_metrics}
\end{figure}

Then, we show the denoising performance of the intermediate variable generated by the proposed method with respect to the iteration numbers in a way similar to \cite{chen2017denoising}.
In particular, we plot the MPSNR, MSSIM, and ERGAS values with respect to the iteration numbers in \cref{fig_conv_metrics}. 
It is seen that with the iteration number increases, the recovered HSI has gradually improved quality in all metrics until convergence.
This implies that the optimization pushes the variable sequence to converge to the optimal solution.
Moreover, the metric values converge within about 20 iterations, which implies that the proposed method generates the desired solution efficiently.

{

\subsection{Ablation Study}
To better show the significance and effectiveness of the log-based nonconvex RPCA approach as well as the exploration of spatial-spectral information in HSI denoising, 
we further conduct some experiments for clear illustration. 
In particular, we conduct experiments using the synthetic data generated in \cref{sec_data_syn} for a quantitative illustration.
We divide the proposed model into two parts, including the nonconvex RPCA and SSTV parts, which correspond to the first and second terms of \cref{eq_obj}, respectively. 

First, we conduct experiments to show the significance of exploiting spatial-spectral information of HSI images in our model. 
For this purpose, we compare the performance of our model under two conditions, i.e., with $\gamma > 0$ or $\gamma = 0$, and show the results in \cref{fig_ablation_sstv}.
It is seen that when $\gamma = 0$, our model falls back to \cref{eq_obj_log}, where the SSTV term is not adopted.
For the two approaches, we report their highest performance by tuning balancing parameters.
It is seen that the proposed model with integrated SSTV term has significantly improved denoising performance than the pure nonconvex RPCA model of \cref{eq_obj_log}.
The only difference between the two models lies in the usage of the SSTV norm and thus it is natural to believe that such difference leads to the difference of denoising performance. 
These observations confirm the significance and effectiveness of exploiting spatial-spectral information by integrating an SSTV term in our model.

\begin{figure}[h]
	\centering
	{\includegraphics[width=0.95\columnwidth]{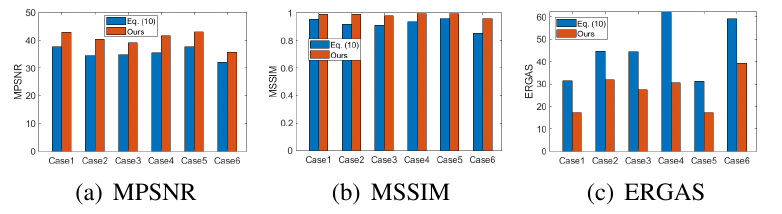} }	
	\caption{ Illustration of the effectiveness of exploiting spatial-spectral information for HSI denoising. }
	\label{fig_ablation_sstv}
\end{figure}
\begin{figure}[h]
	\centering
	{\includegraphics[width=0.95\columnwidth]{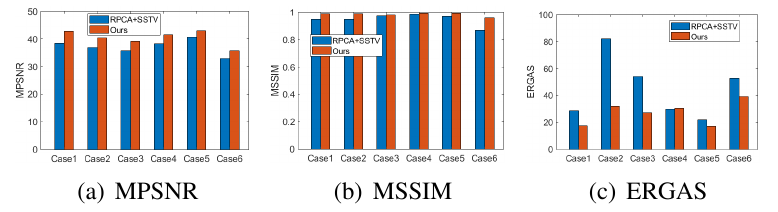} }	
	\caption{ Illustration of the effectiveness of adopting nonconvex RPCA approach for HSI denoising. }
	\label{fig_ablation_rpca}
\end{figure}

Then, we conduct experiments to show the significance of the nonconvex RPCA part in our model. 
For this purpose, we compare our model with ``RPCA+SSTV'', which refers to the model of convex RPCA with an integrated SSTV term. 
We show the comparison results in \cref{fig_ablation_rpca}, where the highest denoising performance is reported for each method, respectively.
We show the comparison results of the two approaches in \cref{fig_ablation_rpca}.
It is seen that our method has significantly improved performance compared with the RPCA+SSTV approach, which suggests the effectiveness of using nonconvex approach in HSI denoising application.

}

\section{Conclusion}
\label{sec_conclusion}
In this paper, we propose a novel method for HSI denoising, named L$^3$S$^3$TV.
Unlike existing low-rank models that only focus on developing more accurate rank approximation for low-rank component recovery,
we propose to simultaneously adopt nonconvex approximations to both the rank and the column-wise sparsity for more accurate separation of the low-rank and sparse components.
In particular, we propose log-based column-wisely sparse approximation, named the $\ell_{2,\log}$ norm, 
which is more accurate than the widely used convex approach, i.e., $\ell_{2,1}$ norm.
For its associated shrinkage problem, we developed an efficient optimization strategy which is formally presented in a theorem.
The $\ell_{2,\log}$ norm can be generally used in various problems that restrict column-wise sparsity. 
Moreover, we impose the SSTV regularization in the log-based nonconvex RPCA model, which enhances the global piece-wise smoothness and spectral consistency from the spatial and spectral views in the recovered HSI.
Extensive experiments on both simulated and real HSIs demonstrate the effectiveness of the proposed method in denoising HSIs.

\section*{Acknowledgment}
 This work is supported by National Natural Foundation of China (NSFC) under Grants 61806106, 61802215, and 61806045, and Natural Science Foundation of Shandong Province under Grants ZR2019QF009 and ZR2019BF011; Q.C. is partially supported by NIH R21AG070909, UH3 NS100606-03 and a grant from the University of Kentucky.

{\appendix	
 In particular, we consider two distributions that centralized data usually follow in most cases.
First, suppose that $X_1,\cdots,X_d\overset{i.i.d}\sim\mathcal{N}(0,1)$, then it is clear that 
$X_1^2+\cdots+X_d^2=\sum_{i=1}^{d}X_i^2\sim\chi^2(d).$ Thus, we have the following bound:
\begin{equation}
\begin{aligned}
& \mathbf{E} \Bigg(\log \Bigg(1 + \sqrt{\sum\nolimits_{i=1}^d X_{i}^2}\Bigg) \Bigg) \\
= & \int_{0}^{+\infty} \log  (1 + \sqrt{ y } ) \frac{1}{2^{\frac{d}{2}} \Gamma(\frac{d}{2})} e^{-\frac{y}{2}} y^{\frac{d}{2} - 1} \dd{y}\\
< & \int_{0}^{\infty} y^{\frac{1}{4}} \frac{1}{2^{\frac{d}{2}} \Gamma(\frac{d}{2})} e^{-\frac{y}{2}} y^{\frac{d}{2} - 1} \dd{y}\\
= & 2^{\frac{1}{4}} \int_{0}^{\infty} \frac{1}{\Gamma(\frac{d}{2})} e^{-\frac{y}{2}} \Big(\frac{y}{2}\Big)^{\frac{d}{2}-\frac{3}{4}} \dd{\frac{y}{2}} \\
= & \frac{\sqrt[4]{2}}{\Gamma(\frac{d}{2})} \int_{0}^{\infty} e^{-\frac{y}{2}} \Big(\frac{y}{2}\Big)^{\frac{d}{2}-\frac{3}{4}} \dd{\frac{y}{2}}  \\
= & \frac{\sqrt[4]{2}\Gamma(\frac{d}{2}+\frac{1}{4})}{\Gamma(\frac{d}{2})}, \\
\end{aligned}
\end{equation}
where $\Gamma(y) = \int_{0}^{+\infty}t^{y-1}e^{-t} \dd{t}  \space (y>0)$ is the Gamma function. 
Second, suppose that $X_1,\cdots,X_d\overset{i.i.d}\sim\mathcal{U}(0,1)$, 
then we have:
\begin{equation}
\begin{aligned}
&	\mathbf{E}(X_1,X_2,\cdots,X_d) \\
= & \int\!\cdots\! \int_{-\infty}^{+\infty} \log \Bigg(1 + \sqrt{\sum\nolimits_{i=1}^d x_{i}^2}\Bigg)f(x_1,\cdots,x_d)  \dd{x_1}  \cdots  \dd{x_d}  \\
= & \int\!\cdots\! \int_{0}^{1} \log \Bigg(1 + \sqrt{\sum\nolimits_{i=1}^d x_{i}^2}\Bigg) \prod\nolimits_{j=1}^{d} f_{X_j}(x_j)  \dd{x_1}  \cdots  \dd{x_d} \\
< & \int\!\cdots\! \int_{0}^{1} \sqrt{\sum\nolimits_{i=1}^d x_{i}^2}  \dd{x_1}  \cdots  \dd{x_d}  \\
\le & \int\!\cdots\! \int_{0}^{1} \sum\nolimits_{i=1}^d x_{i}  \dd{x_1}  \cdots  \dd{x_d} = \frac{d}{2}, \\
\end{aligned}
\end{equation}
where $f(x_1,\cdots,x_d)$ is the joint probability density function and $f_{X_j}(x_j)$ is the probability density function of $X_i$.
}

\bibliographystyle{IEEEtrans}
\bibliography{main}  

\end{document}